\documentclass[twoside,11pt]{article}
\usepackage{subcaption}
\usepackage[preprint]{jmlr2e-arxiv}
\usepackage[utf8]{inputenc}
\usepackage{dsfont}
\usepackage{natbib}
\usepackage{graphicx,graphics}
\usepackage{wrapfig}
\usepackage{amsmath}
\usepackage{enumitem}
\usepackage{booktabs} 
\usepackage{nicefrac} 
\usepackage{yfonts}
\usepackage[euler]{textgreek}
\usepackage[normalem]{ulem}
\usepackage{listings}
\usepackage{MnSymbol}
\usepackage{silence}
\makeatletter
\def\thm@space@setup{%
  \thm@preskip=\parskip \thm@postskip=0pt
}
\makeatother

\usepackage[linesnumbered, ruled, vlined]{algorithm2e}%
\usepackage{algpseudocode}

\def\P{\mathbb{P}}
\def\E{\mathbb{E}}
\def\R{\mathbb{R}}

\newcommand{\argmin}{\mathop{\mathrm{argmin}}}

\usepackage{xcolor}
\usepackage[textsize=tiny]{todonotes}

\newcounter{counter}[section]

\newcommand{\change}[1]{{\leavevmode\color{black}{#1}}}

\begin{document}

\title{Online Auction Design Using Distribution-Free Uncertainty Quantification with Applications to E-Commerce}

\vspace{0.2in}

\author{\name Jiale Han \email jialehan@ucla.edu\\
\addr Department of Statistics and Data Science\\
University of California, Los Angeles, CA 90095-1554, USA\\
\AND
\name Xiaowu Dai\footnotemark[1] \email daix@ucla.edu\\
\addr Department of Statistics and Data Science and Department of Biostatistics\\
University of California, Los Angeles, CA 90095-1554, USA}

\maketitle
\renewcommand{\thefootnote}{\fnsymbol{footnote}}
\footnotetext [1] {\textit{Address for correspondence:} Xiaowu Dai, Department of Statistics, University of California, Los Angeles, 8125 Math Sciences Bldg \#951554,  Los Angeles, CA 90095, USA. Email: daix@ucla.edu.}

\vspace{0.01in}
\begin{center}
To appear in \emph{Journal of the American Statistical Association.}
\end{center}

\vspace{0.15in}
\begin{abstract}
\noindent
Online auction is a cornerstone of e-commerce, and a key challenge is designing incentive-compatible mechanisms that maximize expected revenue. Existing approaches often assume known bidder value distributions and fixed sets of bidders and items, but these assumptions rarely hold in real-world settings where bidder values are unknown, and the number of future participants is uncertain. In this paper, we introduce the Conformal Online Auction Design (COAD), a novel mechanism that maximizes revenue by quantifying uncertainty in bidder values without relying on known distributions. COAD incorporates both bidder and item features, using historical data to design an incentive-compatible mechanism for online auctions.
Unlike traditional methods, COAD leverages distribution-free uncertainty quantification techniques and integrates machine learning methods, such as random forests, kernel methods, and deep neural networks, to predict bidder values while ensuring revenue guarantees.
Moreover, COAD introduces bidder-specific reserve prices, based on the lower confidence bounds of bidder valuations, contrasting with the single reserve prices commonly used in the literature.
We demonstrate the practical effectiveness of COAD through an application to real-world eBay auction data.  Theoretical results and extensive simulation studies further validate the properties of our approach.
\end{abstract}
\bigskip

\noindent%
{\it Keywords:}  Conformal prediction; Online auction; Revenue maximization;  Uncertainty quantification.

\newpage
\section{Introduction}\label{sec:intro}
Online auctions have been central to enabling individuals and businesses to benefit from e-commerce, such as merchants selling items on eBay or platforms selling advertising slots.
For example, in an eBay auction, a seller lists an item, and the winner is determined through an auction. 
In online advertising, advertisers bid for ad slots in real-time auctions on platforms like Google and Meta, where the winning bidder pays the platform to display their advertisement.
Because these major platforms collect extensive data on user behavior, online selling and advertising allow for personalized recommendations and more precise user targeting compared to traditional offline sales or printed advertisements. Online auctions have generated a significant portion of revenue for these platforms \citep{choi2020online}. As a result, the study of online auctions has become a central topic in both computer science and economics. Existing research focuses on analyzing the economic properties of auction designs as well as their computational efficiency \citep{riley1981optimal,  roughgarden2010algorithmic, easley2010networks, ostrovsky2023reserve}.

It is critical to design auctions to achieve incentive compatibility and maximize revenue \citep{milgrom2017discovering}. \citet{myerson1981optimal} developed an auction mechanism by combining the Vickrey–Clarke–Groves (VCG) mechanism \citep{vickrey1961counterspeculation, clarke1971multipart, groves1973incentives} with a reserve price when bidder values are independently drawn from a known regular distribution. 
This pioneering work has laid the foundation for most online auctions for advertisements \citep{easley2010networks}. In these auctions with a single auction item, the highest bidder pays the higher of the second-highest bid or the reserve price set by the platform. This mechanism guarantees incentive compatibility, meaning that bidders are incentivized to bid their true value. However, implementing this mechanism in practice has two main challenges: (i) value distributions are often unknown in real-world auctions, and (ii) setting an appropriate reserve price is difficult in online settings with uncertain, heterogeneous future bidders. 
The central question is how to design an incentive-compatible mechanism that maximizes the platform's revenue in such online scenarios, where future bidders have unknown value distributions.

We propose the Conformal Online Auction Design (COAD), a new mechanism for maximizing revenue by quantifying uncertainty in bidders' valuations without requiring knowledge of the value distribution. 
COAD integrates three key components. First, it uses historical data on both item and bidder features to estimate each bidder's valuation for a new item, along with an uncertainty measure. For instance, in online advertising platforms like Google, bidders are advertisers, and items are ad slots associated with different keywords. Item features may include keywords, while bidder features may include advertiser ratings, brand identity, and product information. By incorporating both bidder and item features, COAD captures the inherent complexity and heterogeneity of real-world online auctions—where assuming fixed value distributions or identical items across auctions is often unrealistic.

Second, COAD adopts a distribution-free, prediction interval–based framework for estimating bidder values, making it compatible with a broad class of machine learning algorithms, including random forests, kernel methods, and deep neural networks. It leverages conformal prediction with conditional guarantees \citep{gibbs2023conformal}, which produces prediction intervals that achieve exact coverage for each group, regardless of the sample size.
For auctioned items with finitely many types and a specified confidence level, COAD constructs prediction intervals for each bidder’s valuation in new auctions. This allows the mechanism to adapt to any item type and any number of new bidders without increasing computational cost, while still ensuring strong revenue guarantees. In contrast, many existing approaches require the sample size to scale with the number of new bidders \citep{devanur2016sample, guo2019settling}.

Third, COAD introduces bidder-specific reserve prices based on the lower confidence bound of each bidder’s estimated valuation, in contrast to the traditional approach of using a uniform reserve price \citep{cesa2014regret, mohri2016learning, ostrovsky2023reserve}. In single-item auctions, a uniform reserve price is only impactful when the number of bidders is small \citep{ostrovsky2023reserve}. However, bidder-specific reserve prices have been shown to be more effective in practice for maximizing revenue. For example, platforms like Google and Yahoo! implement personalized reserve prices to increase revenue while also promoting higher-quality advertisements \citep{even2008position}, illustrating the practical benefits of this strategy. COAD builds on this idea and offers theoretical guarantees: it is incentive-compatible and ensures that the expected revenue is at least a constant fraction of the optimal revenue.

The COAD mechanism differs from existing approaches that rely on historical data to estimate empirical value distributions when bidder distributions are unknown \citep{cole2014sample, huang2015making, morgenstern2015pseudo, roughgarden2016ironing, guo2019settling}. COAD offers two key advantages over these methods.
By incorporating both item and bidder features, COAD generalizes beyond settings with a fixed item and a fixed set of bidders. This flexibility enables its application to auctions involving entirely new items and new bidders. Moreover, new bidders may have features that differ substantially from those of past participants, leading to different value distributions. This variation poses a challenge for traditional methods, which typically require sufficient samples to estimate bidder-specific distributions. In contrast, COAD only requires the features of new bidders to be known in order to construct bidder-specific reserve prices.

The COAD mechanism also differs from existing methods for learning the optimal reserve price in second-price auctions \citep{cesa2014regret, mohri2016learning}. These methods typically rely only on the highest and second-highest bids to estimate a single reserve price. In contrast, COAD makes use of the full bid data, allowing it to extract valuable information even from auctions with just one bidder. This leads to significantly improved data efficiency.
Moreover, COAD does not require the assumption of a known upper bound on bidder values. This makes COAD more flexible for real-world auction settings where bidder valuations can vary widely and such bounds are rarely known.

The remainder of the paper is organized as follows. Section \ref{sec:datadescrip} describes the online auction dataset from eBay. Section \ref{sec:two} introduces the modeling framework. Section \ref{sec:three} presents the proposed COAD mechanism. Section \ref{sec:analysis} provides the theoretical analysis of the mechanism.
Section \ref{sec:real_data} applies COAD to real eBay auction data. Section \ref{sec:four} presents application-based simulation studies. Section \ref{sec:relatedwork} discusses related literature. Section \ref{sec:six} concludes the paper with future directions. Additional simulations and all technical proofs are included in the supplemental materials.

\section{Data Description}
\label{sec:datadescrip}
We study an eBay online auction dataset, which contains 149 seven-day auctions for Palm Pilot M515 PDAs and is publicly available at \url{https://www.modelingonlineauctions.com/datasets}. 
In each auction, bidders placed multiple bids, with each bid exceeding the last, and each bidder's highest bid reflecting their true value for the item. 
We model the historical data by considering each seven-day auction as occurring at a distinct time point, where eBay conducts auctions on behalf of sellers. Multiple bidders enter the auction, each selecting one item to bid on, and those choosing the same item participate in an auction, bidding their true value in the final attempt.
The dataset contains multiple auctions for similar items sold by different sellers, making seller information a feature of the items. Bidders' historical bidding information serves as their features.  Since both sellers and bidders change across auctions, the auction environment is highly heterogeneous.

A central question is how eBay can design an auction mechanism that maximizes sellers’ revenues—thereby attracting more sellers to the platform and increasing overall earnings.  Currently, eBay employs a second-price auction  \citep{cesa2014regret}, which, while incentive-compatible, does not offer strong revenue guarantees.
Many existing methods rely on assumptions such as known bidder value distributions or a fixed set of bidders and items. These assumptions are not valid in the context of online auctions, where future bidders are uncertain and their value distributions are unknown. Moreover, new bidders often have features that differ significantly from those of past participants, resulting in distinct value distributions. This heterogeneity makes it challenging to accurately estimate bidder-specific value distributions for each item \citep{cole2014sample, huang2015making, roughgarden2016ironing}, especially since most bidders appear in only a small number of auctions, leading to limited data availability.

Our goal is to design a mechanism that addresses the following two key questions based on this dataset:
\begin{itemize}
\item[(Q1)] How can we design a mechanism that leverages historical data to guarantee revenue for future auctions, even with limited available data?
\item[(Q2)] How can the designed mechanism handle scenarios involving heterogeneous bidders and items?
\end{itemize}
In the following sections, we use conventional terminology. Specifically, we refer to \emph{auctioneers} as e-commerce or online advertising platforms, such as eBay or Google. The \emph{items} represent goods sold on these platforms, such as Palm Pilot M515 PDAs in the eBay dataset or ad slots on advertising platforms. The \emph{bidders} are those interested in buying the items, like eBay users or advertisers on online platforms.

\section{Online Auction Model with Features}
\label{sec:two}

\begin{figure}[ht]
    \centering
    \includegraphics[width=0.8\textwidth]{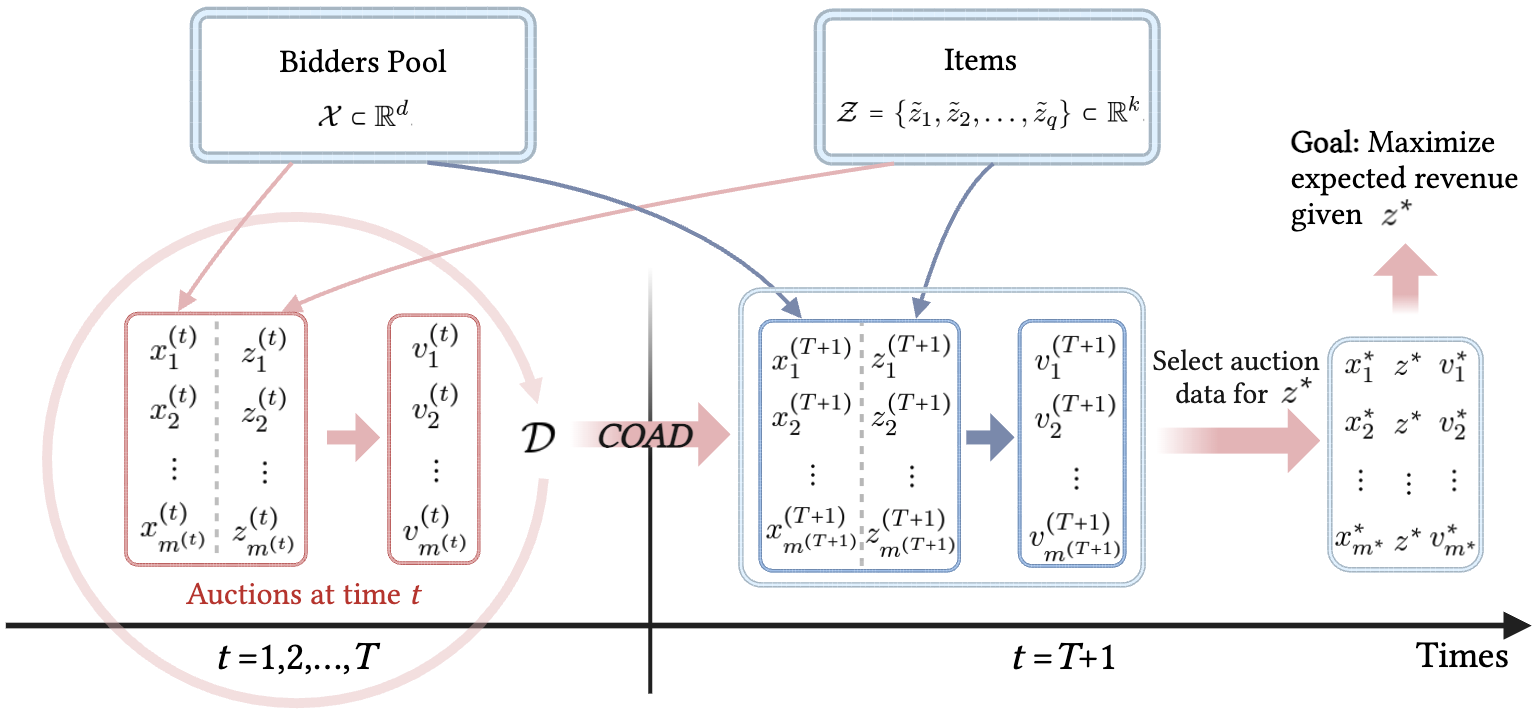}
    \caption{\small{An illustration of the online auction process.}}
    \label{fig:flow}
\end{figure}

\noindent
We begin by describing the online auction model. Consider an auctioneer (e.g., an e-commerce platform) with a finite variety of indivisible items, each available in unlimited supply (e.g., ad slots). 
The auctioneer holds auctions at $T$ time points, each consisting of several single-round auctions for individual items. Each time point involves multiple bidders, each participating in different auctions.
At each time $t=1,\dots, T$, a bidder can participate in only one auction, and each item is sold only once. Each item has a feature from a finite set $\mathcal{Z}=\{\tilde{z}_1,\tilde{z}_2, \dots,\tilde{z}_{q}\}$, where each $\tilde{z}_j\in\mathbb{R}^k$ for $1\leq j\leq q$. Suppose there are $m^{(t)}\geq 1$ bidders at time $t$, denoted as $[m^{(t)}]=\{1,2,\dots, m^{(t)}\}$. Each bidder $j\in[m^{(t)}]$ has a feature $x^{(t)}_j\in \mathcal{X}\subset\mathbb{R}^d$, participates in the auction for an item with feature $z_j^{(t)}\in\mathcal{Z}$, and has a valuation for the item, $v_j^{(t)}\in\mathbb{R}_{\geq 0}$.
This setting aligns with our description of the eBay dataset in Section \ref{sec:datadescrip} and is also widely applicable to other online auctions. For instance, in online advertising, the items are ad slots associated with different keywords, where item features could include information related to these keywords. Advertisers bid for ad slots linked to specific keywords, selecting one keyword and bidding once per session. Advertiser features might include rating level, ad brand, product information, and other details relevant to the advertisement.
The online auction process is shown in Figure \ref{fig:flow}. 

\subsection{Online Auction Design}\label{sec:online}
In this section, we model the decision-making process and auction design for the new auctions at time $T+1$. Our goal is to design an online auction mechanism for any specific item that not only incentivizes bidders to reveal their true values but also maximizes expected revenue by leveraging historical data up to time $T$, without knowing the bidders' value distributions. 
At time $T+1$, we focus on an auction for an item with feature $z^*$ that is randomly selected from the feature set $\mathcal{Z}$, representing a specific auction at this time point. We assume this auction attracts $m^*$ bidders.
We let $[m^{*}]=\{1,2,\dots, m^{*}\}$ as the set of bidders in the new auction. Consider that each bidder $i\in[m^*]$ with feature $x^*_i\in\mathcal{X}$ and value $v^*_i$ submits a bid $b^*_i\in\mathbb{R}_{\geq 0}$. We define $\vec{v}^*=(v^*_1,v^*_2,\dots, v^*_{m^*})$ as the vector of $m^*$ bidders' values, $\vec{x}^*=(x^*_1,x^*_2,\dots,x^*_{m^*})$ as the vector of their features, and $\vec{b}^*=(b^*_1,b^*_2,\dots, b^*_{m^*})$ as the vector of their bids.

For any $m^*\in \mathbb{N}_{+}$ and $(\vec{b}^*,\vec{x}^*,z^*)\in\mathbb{R}^{m^*}_{\geq 0}\times \mathcal{X}^{m^*}\times \mathcal{Z}$, an auction mechanism is a mapping from the space of  $(\vec{b}^*,\vec{x}^*,z^*)$ to an allocation rule $a_i(\vec{b}^*,\vec{x}^*, z^*)\in [0,1]$ for each bidder $i\in[m^*]$. This allocation rule denotes the probability of bidder $i$ being allocated the item. Additionally, a payment rule, denoted by $p_i(\vec{b}^*,\vec{x}^*,z^*)\in \mathbb{R}_{\geq 0}$, specifies the price paid by bidder $i$. 
Since only one item is to be allocated, the allocation rule satisfies,
\begin{equation}\label{eq1}
\sum_{i=1}^{m^*}a_i(\vec{b}^*,\vec{x}^*,z^*)\leq 1 \quad \text{and} \quad a_i(\vec{b}^*,\vec{x}^*,z^*)\geq 0, 
\end{equation}
for any $i\in[m^*]$ and $(\vec{b}^*,\vec{x}^*,z^*)\in\mathbb{R}^{m^*}_{\geq 0}\times \mathcal{X}^{m^*}\times\mathcal{Z}$. 
In this paper, we focus on a deterministic mechanism so that $a_i\in\{0,1\}$. We assume that the bidders behave myopically, optimizing their utility based solely on the current auction. Specifically, the utility of bidder $i$ is defined as $ u_i(\vec{b}^*,\vec{x}^*,z^*) = v^*_i \cdot a_i(\vec{b}^*, \vec{x}^*, z^*) - p_i(\vec{b}^*, \vec{x}^*, z^*) $. 

Since each bidder's valuation is private information, a bidder might not disclose their true value while bidding but may submit a different value to game the mechanism and attempt to achieve higher utility. Therefore, a primary objective in mechanism design for auctions is to incentivize bidders to bid their true values. The auctioneer aims to design a mechanism that maximizes expected revenue,
subject to the constraints of \emph{dominant strategy incentive compatibility (IC)} given by,
\begin{equation}\label{eqn:ic}
    u_i(v^*_i, \vec{b}^*_{-i},\vec{x}^*,z^*)\geq  u_i(b^*_i, \vec{b}^*_{-i},\vec{x}^*,z^*),
\end{equation}
for every $ v^*_i,b^*_i\in \mathbb{R}_{\geq 0}$ and $(\vec{b}^*_{-i}, \vec{x}^*, z^*)\in\mathbb{R}^{m^*-1}_{\geq 0}\times\mathcal{X}^{m^*}\times\mathcal{Z}$, and \emph{individual rationality (IR)},
\begin{equation}\label{eqn:ir}
    u_i(v^*_i, \vec{b}^*_{-i},\vec{x}^*,z^*)\geq 0.
\end{equation}
Under IC and IR, the dominant strategy for the bidders to obtain the highest utility is to bid truthfully, that is, $\vec{b}^*=\vec{v}^*$. 

\subsection{Learning with Bidder and Item Features}\label{sec:feature}
When there is no information about the values $\vec{v}^*$, any deterministic auction mechanism that satisfies IC in \eqref{eqn:ic} and IR in \eqref{eqn:ir}  can perform arbitrarily poorly in terms of revenue \citep[e.g.,][]{sandholm2015automated}. Hence, we consider a setting in which, although the valuations, prior distributions of values, and support information of values are all \emph{unknown} to the auctioneer, the auctioneer has access to historical data from the previous auctions up to time $T$. 
Denote the data of historical auctions as $\mathcal{D}=\{(x_j^{(t)},z_j^{(t)}, v_j^{(t)})  \ |\  j\in[m^{(t)}], \ t=1,2\dots,T\}$, which includes bidder features $x_i^{(t)}$, bidder values $v_i^{(t)}$, and item features $z_i^{(t)}$ from all auctions up to time $T$.
For any $(x,z,v)\sim P$, we consider the following regression model for the remainder of this paper:
\begin{equation}\label{model_1}
     v=\mu(x,z)+\varepsilon,
\end{equation}
where $\mu(x,z)=\mathbb{E}[v|x,z]$ represents the expected effect of bidder and item features on bidder values. 
Let $\sum_{t=1}^{T}m^{(t)}=N$. We rewrite $\mathcal{D}=\{(x_j,z_j,v_j)  \ |\  j=1,2,\dots, N\}$ for notation simplicity and make the following three assumptions. 
\begin{assumption}[iid historical data]
\label{assump:iiddata}
The $N$ data points in $\mathcal{D}$ are independent and identically distributed (iid) copies of $(x,z,v)\sim P$, where $P$ is a distribution function on $\mathcal{X}\times\mathcal{Z}\times\mathbb{R}_{\geq 0}$.
\end{assumption}
\begin{assumption}[Bounded error]
\label{assump:indepnoise}  
For any \((x, z, v) \sim P\), the value approximation error \(|\epsilon|=|v - \mathbb{E}[v| x, z]|\) is bounded by
$
C := \sup_{(x, z, v) \sim P} |v - \mu(x, z)|<\infty.
$
\end{assumption}
\begin{assumption}[Independent new bidders]
\label{assump:indepbidders}
The pairs $(x^*_i,v^*_i), 1\leq i\leq m^*$, are iid conditional on $z^*$. Additionally, 
$\{(x^*_i,z^*,v^*_i), (x_1,z_1,v_1),\dots, (x_{N},z_{N},v_{N})\}  \stackrel{\text{iid}}{\sim} P$,  $\forall i\in[m^*]$.
\end{assumption}
\noindent
Assumption \ref{assump:iiddata} is standard in supervised learning and widely used in auction literature \citep{mohri2016learning, lei2018distribution,ostrovsky2023reserve}. 
Assumption \ref{assump:indepnoise} is relaxed than assuming a known upper bound on bidders' values \citep[e.g.,][]{babaioff2017posting}. For example, if \( v \) is constrained within \([1,h]\), then \( C \leq h - 1 \) always holds.
For each $i \in [m^*]$, let $\varepsilon_i^*=v_i^*-\mu(x^*_i,v^*_i)$. 
Assumption \ref{assump:indepbidders} considers bidders in a new auction for an item with feature $z^*$ at time $T+1$ are independent. Additionally, the data $(x^*_i,z^*,v^*_i)$ is assumed to be iid relative to the historical data.
This assumption allows for different value distributions among bidders, depending on their features. Specifically,  $(x^*_i,v^*_i)|z^*$ are iid, but when both bidder and item features are considered, the distribution of $v_i^*|(x_i^*,z^*)$ varies, as $v_i^*=\mu(x_i^*,z^*)+\varepsilon_i^*$, with the mean changing due to differences in $x_i^*$. 

We introduce additional notations. For any $(x^*,z^*,v^*)\sim P$, let  $F_{v^*|z^*}$ and $F_{v^*, x^*|z^*}$ be the distribution functions for $v^*|z^*$ and $(v^*,x^*)|z^*$, respectively. Let $F_{\vec{v}^* |z^*}$ and  $F_{\vec{v}^*, \vec{x}^*|z^*}$ be the joint distributions for the vectors $\vec{v}^* |z^*$ and $(\vec{v}^*, \vec{x}^*)|z^*$, respectively.  Under Assumption \ref{assump:indepbidders}, $F_{\vec{v}^*|z^*}(\vec{v}^*)= \prod_{i=1}^{m^*}F_{v^*|z^*}(v_i^*)$ and $F_{\vec{v}^*, \vec{x}^*|z^*}(\vec{v}^*,\vec{x}^*)=\prod_{i=1}^{m^*}F_{{v^*},{x^*}|z^*}({v}_i^*,{x}_i^*)$.

\subsection{Revenue Objective}
\label{sec:revenueobjective}
For an auction mechanism $\mathcal{M}$ designed using historical data $\mathcal{D}$, let $R_{m^*}^{\mathcal M|\mathcal{D}}(F_{{v^*}, {x^*}|z^*})$ denote the expected revenue generated by $\mathcal{M}$ when there are $m^*$ random bidders in the auction for an item with feature $z^*$. That is,
\begin{equation}\label{rm}
    R_{m^*}^{\mathcal M|\mathcal{D}}(F_{{v^*}, {x^*}|z^*})= \mathbb{E}\left[\sum^{m^*}_{i=1} p_i(\vec{v}^*,\vec{x}^*,z^*) \big| \mathcal{D}, z^*\right].
\end{equation}
Here $p_i(\vec{b}^*,\vec{x}^*,z^*)$ is the price paid by bidder $i\in[m^*]$, and the expectation in \eqref{rm} is taken over the distribution $F_{\vec{v}^*, \vec{x}^*|z^*}$. 
Let $W_{m^*}(F_{{v}^*|z^*})$ be the maximum expected social welfare of the $m^*$ random bidders,
\begin{equation}\label{sw}
    W_{m^*}(F_{{v^*}|z^*}) = \mathbb{E}_{}\left[\max_{1\leq i\leq m^*} v_i^*  \big|  z^*\right],
\end{equation}
where the expectation in \eqref{sw} is taken over the distribution $F_{\vec{v}^*|z^*}$. 
Since for a given item, the payments of the bidders depend on both the features and the values of the bidders, while the maximum value among the bidders depends only on the values of the bidders, there is a difference between the underlying distributions in $ R_{m^*}^{\mathcal M|\mathcal{D}}$ and $W_{m^*}$.

The maximum expected social welfare in \eqref{sw} is a benchmark for revenue in \eqref{rm} because it can be shown, using equations  \eqref{eq1} and \eqref{eqn:ir},  that for any mechanism $\mathcal M$, $\sum^{m^*}_{i=1} p_i(\vec{v}^*,\vec{x}^*,z^*)\leq \max_{1\leq i\leq m^*}v_i^*$ holds for all $(\vec{v}^*,\vec{x}^*,z^*)\in\mathbb{R}^{m^*}_{\geq 0}\times \mathcal{X}^{m^*}\times\mathcal{Z}$. As a result,   $R^{\mathcal M|\mathcal{D}}_{m^*}(F_{{v^*}, {x^*}|z^*})\leq W_{m^*}(F_{{v^*}|z^*})$ holds for any mechanism $\mathcal M$ and distribution $P$,  indicating that  $W_{m^*}(F_{{v^*}|z^*})$ represents the optimal achievable revenue. However, achieving the revenue of $W_{m^*}(F_{{v^*}|z^*})$  would require the auctioneer to know the exact values  $\vec{v}^*$ for the item with feature $z^*$ before the auction, which is impractical in real-world scenarios.

\section{Conformal Online Auction Design}\label{sec:three}
In this section, we present a new algorithm called \emph{conformal online auction design} (COAD). The key idea is to employ prediction intervals with high-confidence coverage for bidders' values of any given item, which we derive using conformal prediction with a conditional guarantee. We also demonstrate the statistical efficiency of the conditional conformal prediction method employed in our algorithm.

\subsection{The COAD Mechanism}\label{mechanism}

\begin{algorithm}[t!]
\caption{ \normalsize{{Conformal Online Auction Design (COAD)}}}
\begin{algorithmic}[1]
\State  \normalsize{\textbf{Input:} Historical data $\mathcal{D}=\{(x_i,z_i,v_i) \ |\  i=1,2,\dots, N\}$; New auction data $\{(x_i^*,z^*,v_i^*) \ |\  i=1,2,\dots, m^*\}$ at time $T+1$; Miscoverage level $\alpha\in(0,1)$.  }
\State \textbf{for} $i=1$ to $m^*$ \textbf{do}
\State \quad \textbf{Step 1:}  Construct the $(1-\alpha)$-prediction interval $[\hat{v}^L_i,\hat{v}^U_i]$ for $v_i^*$ by  Eq. \eqref{dual_new};
\State \quad \textbf{Step 2:} Obtain the pseudo-virtual values $c_i(v_i^*,x_i^*,z^*)$ by Eq. \eqref{eqn:virvalue};
\State \textbf{end for}
\State \textbf{for} $i=1$ to $m^*$ \textbf{do}
\State \quad \textbf{Step 3:}    Determine the allocation rule $a_i(\vec{v}^*,\vec{x}^*,z^*)$ by the following procedure:
\State \quad\quad \textbf{if} $\max_{i\in [m^*]}c_i(v_i^*,x_i^*,z^*)=0$ \textbf{then}  $a_i(\vec{v}^*,\vec{x}^*,z^*)=0$;
\State  \quad\quad \textbf{else if} $c_i(v_i^*,x_i^*,z^*)=\max_{i\in [m^*]}c_i(v_i^*,x_i^*,z^*)$ \textbf{then} $a_i(\vec{v}^*,\vec{x}^*,z^*)=1$;
\State \quad\quad   \textbf{else} $a_i(\vec{v}^*,\vec{x}^*,z^*)=0$;
\State \quad\quad \textbf{end if}
\State \quad \textbf{Step 4:} Calculate payment $p_i(\vec{v}^*,\vec{x}^*,z^*)$ by Eq. \eqref{eq:pay}.
\State \textbf{end for} 
\State \textbf{Output:} The allocations $\{a_i(\vec{v}^*,\vec{x}^*,z^*)\}_{i=1}^{m^*}$ and payments $\{p_i(\vec{v}^*,\vec{x}^*,z^*)\}_{i=1}^{m^*}$ for all bidders in $[m^*]$.
\end{algorithmic}
\label{alg:coad}
\end{algorithm}

\noindent
Our conformal online auction design (COAD) consists of four main steps. 
The first step is to construct high-confidence coverage prediction intervals for the bidders' values for any given item. Specifically, for each bidder $i\in[m^*]$, we aim to construct a $(1-\alpha)$-prediction interval $[\hat{v}^L_i,\hat{v}^U_i]$ for their value $v_i^*$, given the item's feature $z^*\in\mathcal{Z}$.
That is,
$\mathbb{P}(v_i^* \in [\hat{v}^L_i,\hat{v}^U_i]\ |\ z^*=\tilde{z}) \geq 1-\alpha$, for any $\tilde{z}\in\mathcal{Z}$ and $\alpha\in(0,1)$, 
where $\hat{v}^L_i$ and $\hat{v}^U_i$ are functions of $x_i^*$ and $z^*$. Although these prediction intervals convey less information compared to the full distribution of values, they are more readily obtainable. The method for constructing these prediction intervals $[\hat{v}^L_i,\hat{v}^U_i]$ is presented in Section \ref{sec:conformal}.

In the second step, we introduce the concept of the \emph{pseudo-virtual value}, which recalibrates each bid to maximize the auctioneer's expected revenue. It is defined as,
\begin{equation}
\label{eqn:virvalue}
    c_i(v_i^*,x_i^*,z^*)=v_i^*\mathbb{I}\{v_i^*\geq \hat{v}_i^L\},\quad\forall i\in[m^*],
\end{equation}
where $c_i$ is a monotone increasing function of $v_i^*$.  Unlike the classical \emph{virtual value} in optimal auction theory \citep{myerson1981optimal}, which depends on the bidder's own valuation and the distribution of valuations among all bidders, the pseudo-virtual value in \eqref{eqn:virvalue} recalibrates bids based on their comparison to the lower bound of the prediction interval. In the classical setting, virtual values adjust bidder valuations according to the probability of higher bids from competitors. However, since the distribution of valuations is unknown in our setting, we instead use the lower confidence bound to establish a competitive threshold for each bid. This approach effectively introduces competition between weaker bidders (those with lower valuations) and stronger bidders, enabling the auctioneer to extract more revenue from the latter. By using this mechanism, the auctioneer can set a high selling price for strong bidders while ensuring that they remain truthful, even when they are aware that their valuations exceed those of all other bidders.

In the third step, we determine the allocation rules $\{a_i(\vec{v}^*,\vec{x}^*,z^*)\}_{i=1}^{m^*}$ as described in \eqref{eq1}. 
We employ a deterministic mechanism where only one bidder can win the item, meaning that at most one value in the set $\{a_i(\vec{v}^*,\vec{x}^*,z^*)\}_{i=1}^{m^*}$ is $1$, with all others set to $0$.
We let the auctioneer retain the item at the new auction if $\max_{i\in [m^*]}c_i(v_i^*,x_i^*,z^*)=0$,
or assigns it to the bidder with the highest pseudo-virtual value otherwise.
If the auctioneer retains the item, then for any bidder $i\in[m^*]$, $ a_i(\vec{v}^*,\vec{x}^*,z^*)=0$.
If there is a tie between the bidders' pseudo-virtual values when $\max_{k\in [m^*]}c_k(v_k^*,x_k^*,z^*)> 0$, for example, 
$    c_i(v_i^*,x_i^*,z^*)=c_j(v_j^*,x_j^*,z^*)=\max_{k\in [m^*]}c_k(v_k^*,x_k^*,z^*)$, 
the auctioneer may break the tie by giving the item to the bidder with the largest prediction lower bound; if there is still a tie, for example, $\hat{v}_i^L=\hat{v}_j^L$, then the auctioneer can break the tie by giving it to the lower-numbered bidder or by using other arbitrary rules. 
After breaking the tie, the winner will have an allocation rule with value $1$, and all others have an allocation rule with value 0.

Finally, the fourth step involves determining the payment. 
In our model, the payment is designed as follows. Let
\begin{equation*}
    r_i(\vec{v}^*_{-i},\vec{x}^*,z^*)=\text{inf}\{b_i^*\ |\ c_i(b_i^*,x_i^*,z^*)\geq 0, c_i(b_i^*,x_i^*,z^*)\geq c_j(v_j^*,x_j^*,z^*),  \forall j\in[m^*], j\neq i\},
\end{equation*}
which represents the lowest winning bid for bidder $i$ against values of other bidders $\vec{v}^*_{-i}$. Then the payment is defined by
\begin{equation}\label{eq:pay}
p_i(\vec{v}^*,\vec{x}^*,z^*)= \begin{cases}
r_i(\vec{v}_{-i}^*,\vec{x}^*,z^*),\quad a_i(\vec{v}^*,\vec{x}^*,z^*)=1, \\
0,\quad\quad\quad\quad\quad\quad\    a_i(\vec{v}^*,\vec{x}^*,z^*)=0,
\end{cases}
\end{equation}
where $i\in[m^*],  \vec{v}^*\in\mathbb{R}^{m^*}_{\geq 0}, (\vec{x}^*,z^*)\in \mathcal{X}^{m^*}\times\mathcal{Z}$. We show in Section \ref{sec:icandir} that the payment structure in \eqref{eq:pay} ensures the COAD mechanism is IC and IR for the new auction at time $T+1$. The four-step COAD procedure is summarized in Algorithm \ref{alg:coad}.

\subsection{Construction of Prediction Intervals} \label{sec:conformal}
We now construct the prediction intervals in Section \ref{mechanism}  based on the historical data $\mathcal{D}$.
Without loss of generality, let the number of data points $N=2n$.
We randomly split the $2n$ data points equally into two sets: a set of \emph{training data}, and a set of \emph{calibration data}. To simplify the notations, we let $D_{\text{cal}}=\big\{(x_j,z_j,v_j)\ |\ j=1,2,\dots, n\big\}$ denote the set of calibration data, and $D_{\text{train}}=\big\{(x_j,z_j,v_j)\ |\ j=n+1,n+2,\dots, 2n\big\}$ denote the set of training data. 
We can use the machine learning algorithms $\mathcal{A}_n$ to estimate the regression function $\mu$ in Assumption \ref{assump:iiddata} based on the set of training data. That is, $\hat{\mu}_n=\mathcal{A}_n(D_{\text{train}})$. A recently developed conformal prediction method constructs prediction intervals for new response variables, offering conditional guarantees \citep{gibbs2023conformal} that are particularly pertinent to our analysis. 

We outline the key steps of the conditional conformal prediction method, with additional details provided in supplemental materials. For each bidder $i\in[m^*]$ in the new auction for an item with feature $z^*$ at time $T+1$, we define a primal prediction interval for their value $v_i^*$ as follows,
\begin{equation}
\label{eqn:primalconformal}
    \hat{\mathcal{C}}_{\text{primal}}(x_i^*,z^*) = \{v: \mathcal{S}(\{x_i^*,z^*\},v) \leq \hat{g}^i_{\mathcal{S}(\{x_i^*,z^*\},v)}(x_i^*,z^*)\},
\end{equation}
where $\mathcal{S}: \mathcal{X}\times \mathcal{Z} \times \mathbb{R}_{\geq 0}\to \R$ is the conformity score function, defined as $\mathcal{S}(\{x,z\},v) = |v - \hat{\mu}_n(x,z)|$ for $(x,z,v)\in \mathcal{X}\times \mathcal{Z} \times \mathbb{R}_{\geq 0}$.
Given a miscoverage level $\alpha\in(0,1)$, the function $\hat{g}^i_{\mathcal{S}(\{x_i^*,z^*\},v)}: \mathcal{X}\times \mathcal{Z}  \to \mathbb{R}$ depends on both the calibration dataset $D_{\text{cal}}$ and the training dataset $D_{\text{train}}$, ensuring valid coverage guarantees for $v_i^*$ conditional on $z^*$.  While the primal prediction interval provides accurate coverage guarantees, it is computationally expensive and does not yield an explicit closed-form interval. 

To efficiently compute the interval, we transition to using the dual form of the prediction interval for the value $v_i^*$. 
The dual prediction interval can be written as,
\begin{equation}\label{dual_new}
\begin{aligned}
    \hat{\mathcal{C}}_{\text{dual}}(x_i^*,z^*) = [\hat{v}^L_i, \hat{v}^U_i]  =  [\hat{\mu}_n(x_i^*,z^*) - S^*, \hat{\mu}_n(x_i^*,z^*) + S^*], \quad \forall i\in[m^*].
\end{aligned}
\end{equation}
Here, $S^*>0$ is chosen to ensure that $\mathbb{P}(v_i^*\in \hat{\mathcal{C}}_{\text{dual}}(x_i^*,z^*)\ |\ z^*=\tilde{z})\geq 1-\alpha, \forall \tilde{z}\in \mathcal{Z}$. At a high level, the dual prediction interval $ \hat{\mathcal{C}}_{\text{dual}}(x_i^*,z^*)$ is derived from the primal interval $\hat{\mathcal{C}}_{\text{primal}}(x_i^*,z^*)$ by removing a negligible portion at the boundary, where the conformity score function exactly matches its threshold: $\{v: \mathcal{S}(\{x_i^*,z^*\},v) = \hat{g}^i_{\mathcal{S}(\{x_i^*,z^*\},v)}(x_i^*,z^*)\}$. Due to its computational efficiency, we adopt the dual prediction interval in the proposed COAD framework in Algorithm \ref{alg:coad}.

\subsection{Properties of the Conformal Prediction Intervals}\label{sec:propconformal}

We establish an upper bound on the length of the dual conformal prediction interval in \eqref{dual_new}. For any $x,z\in \mathcal{X}\times\mathcal{Z}$, we define the prediction error as $\Delta_n(x,z)=\hat{\mu}_n(x,z)-\mu(x,z)$.

\begin{proposition}\label{prop:length}
Under Assumptions \ref{assump:iiddata}–\ref{assump:indepbidders}, the half-length  of the dual prediction interval \( S^* \) in \eqref{dual_new} satisfies  
$S^* \leq \max_{1 \leq j \leq n} |\Delta_n(x_j,z_j)|  + C$,
where \( C \) is defined in Assumption \ref{assump:indepnoise}.
\end{proposition}

\noindent
Proposition \ref{prop:length} shows that the half-length of the prediction interval is bounded by the maximum prediction error in the calibration data plus the upper bound of the value approximation error $C$. 
A more accurate estimator $\hat{\mu}_n$ results in a smaller $|\Delta_n(x_j,z_j)|$, leading to tighter conformal prediction intervals. In the revenue analysis of COAD, we impose the following mild condition on the estimator $\hat{\mu}_n$. 

\begin{assumption}[Convergence of the estimator]
\label{assump:consistency}
The machine-learning estimator $\hat{\mu}_n$ satisfies that $ \mathbb{E}\left[|\Delta_n(x,z)|^2\right]=O(n^{-2\tau})$, for some $\tau>0$.
\end{assumption}

\noindent
Assumption \ref{assump:consistency} holds for a wide range of popular machine learning methods. For instance, it holds for the $l_1$-penalized linear regression in a variety of sparse models \citep[e.g.,][]{bickel2009simultaneous},  a class of regression trees and random forests \citep[e.g.,][]{wager2015adaptive}, a class of neural networks  \citep[e.g.,][]{chen1999improved}, and a class of kernel methods \citep[e.g.,][]{dai2023orthogonalized}. In each of these methods, Assumption \ref{assump:consistency} is satisfied with $\tau\geq 1/8$.

Finally, we point out that under Assumptions \ref{assump:iiddata} and  \ref{assump:indepbidders},
the  dual prediction interval exhibits the following coverage property \citep[see, e.g.,][]{gibbs2023conformal},
\begin{equation*}
\Big|\P(v_i^*\in \hat{\mathcal{C}}_{\text{dual}}(x_i^*,z^*)\ |\ \mathcal{D}, z^*=\tilde{z})-(1-\alpha) \Big|=O_{\P}\left(\sqrt{\frac{q}{n}}\right).
\end{equation*}
Hence, the dual conformal prediction interval provides exact coverage as 
$n\to\infty$.

\section{Analysis of Incentives and Revenues}\label{sec:analysis}
In this section, we analyze the economic properties of the proposed COAD mechanism in Algorithm \ref{alg:coad}, focusing on incentive guarantees and revenue performance. We also compare COAD with alternative data-driven auction designs.

\subsection{Incentive Guarantees}
\label{sec:icandir}
COAD can be viewed as a second-price auction with bidder-specific reserve prices. As a result, it inherits the incentive properties of the standard second-price auction. The following theorem formally establishes the incentive guarantees for bidders.
\begin{theorem}\label{prop:incentive}
The COAD mechanism enjoys the IC and IR properties as defined in Eqs. \eqref{eqn:ic} and \eqref{eqn:ir}, respectively.
\end{theorem}
\noindent
The proof relies on the well-known envelope formula \citep{myerson1981optimal} and is provided in the supplementary material. 
By Theorem \ref{prop:incentive}, bidders have a dominant strategy to truthfully report their valuations in new auctions at time $T+1$.
\subsection{Revenue Analysis}\label{sec:revenue}
We now analyze the revenue performance of the COAD mechanism. In particular, we show that as the number of bidders $m^*$ increases, the expected revenue also increases.
\begin{theorem}\label{pp2}
    For any given item with feature $z^*=\tilde{z}\in \mathcal{Z}$, 
    the expected revenue of COAD will increase as the number of bidders $m^*$ increases. That is, for any $m_1^*\geq m_2^*\geq 1$,
    \begin{equation*}
       R_{m_1^*}^{\text{COAD}|\mathcal{D}}(F_{{v^*}, {x^*}|z^*=\tilde{z}})\geq R_{m_2^*}^{\text{COAD}|\mathcal{D}}(F_{{v^*}, {x^*}|z^*=\tilde{z}}).
    \end{equation*}
\end{theorem}
\noindent
This result supports an intuitive property of COAD: as more independent bidders participate in the auction, competition increases, leading to higher revenues for the auctioneer. 
This finding aligns with the well-known result in second-price auctions established by \citet{bulow1996auctions}, which shows that adding an additional bidder and setting a zero reserve price is always more beneficial than setting the optimal reserve price.
Both Theorem \ref{pp2} and the results of \citet{bulow1996auctions} emphasize the advantage of attracting more bidders. This result is fundamentally due to the structure of COAD, which can be viewed as a second-price auction with bidder-specific reserve prices.

Next, we establish a lower bound for the expected revenue of the COAD mechanism.

\begin{theorem}\label{prop1}
For any given item with feature $z^*=\tilde{z}\in \mathcal{Z}$ and any $\alpha\in(0,1)$, under Assumptions \ref{assump:iiddata}-\ref{assump:consistency}, 
the expected revenue of the \text{COAD} satisfies
    \begin{equation*}
        R_{m^*}^{{\text{COAD}}|\mathcal{D}}(F_{{v^*}, {x^*}|z^*=\tilde{z}})\geq(1-\alpha)(1-\lambda)W_{m^*}(F_{{v^*}|z^*=\tilde{z}})+O_{\P}\left(n^{-\min\{\tau,1/2\}}\right),
    \end{equation*}
where $\tau$ is defined in Assumption \ref{assump:consistency}, and $\lambda=(C+S^*)/W_{m^*}(F_{{v^*}|z^*=\tilde{z}})$ with $C$ defined in Assumption \ref{assump:indepnoise}. 
\end{theorem}
\noindent
Theorem \ref{prop1} provides a revenue guarantee by comparing the revenue of COAD to the maximum expected social welfare, $W_{m^*}(F_{{v^*}|z^*=\tilde{z}})$, as defined in Section \ref{sec:revenueobjective}. It shows that, for a new auction at $T+1$, the expected revenue of COAD is asymptotically no less than a constant fraction of the maximum expected social welfare. 
The ratio of the expected revenue of COAD to the maximum expected social welfare in the worst-case scenario improves as the number of bidders \( m^* \) increases, since \( \lambda \) decreases with growing \( m^* \).  This result aligns with the findings in \cite{dhangwatnotai2010revenue}, which, under certain assumptions on bidders' value distributions, such as the monotone hazard rate condition, show that their mechanism achieves an expected revenue of at least 
$(m^*-1)/4 m^*$ of the maximum expected social welfare for \(m^* \geq 2\).
Unlike their approach, which assumes all bidders share the same feature, COAD allows for bidders with entirely different features. In addition, the theoretical revenue guarantee for COAD  holds without any assumptions on the distribution of bidders’ values, making it broadly applicable.

\subsection{Optimizing Revenue}\label{sec:sensitivity}
We can maximize the revenue of the COAD mechanism through two main strategies. The first strategy is to increase the number of bidders $m^*$. By Theorem \ref{pp2},  a larger $m^*$ directly leads to higher expected revenue.

The second strategy focuses on improving the accuracy of the estimator \( \hat{\mu}_n \) to reduce prediction error. 
As discussed in Section \ref{sec:revenueobjective},  the maximum expected social welfare $W_{m^*}(F_{{v^*}|z^*})$ serves as the optimal benchmark for maximizing the revenue of COAD. 
Given the lower bound in Theorem \ref{prop1}, our goal is to maximize the ratio \( (1-\alpha)(1-\lambda) \), where \(\lambda\) is defined in Theorem \ref{prop1} as \( \lambda = (C+S^*)/W_{m^*}(F_{{v^*}|z^*=\tilde{z}}) \).
There is generally a trade-off between the miscoverage level \( \alpha \) and the half-length of the prediction interval \( S^* \), as a lower miscoverage level typically results in a wider interval. In the worst-case scenario, Proposition \ref{prop:length} ensures that \( S^* \) is always bounded by the prediction error plus a constant. This leads to the following corollary, which establishes a revenue lower bound for COAD without requiring Assumptions \ref{assump:iiddata} and \ref{assump:indepbidders}.
\begin{corollary}\label{cor 1}
    For any given item with feature $z^*=\tilde{z}\in \mathcal{Z}$, under Assumptions \ref{assump:indepnoise} and \ref{assump:consistency},  
the expected revenue of COAD satisfies
    \begin{equation*}
        R_{m^*}^{{\text{COAD}}|\mathcal{D}}(F_{{v^*}, {x^*}|z^*=\tilde{z}})\geq W_{m^*}(F_{{v^*}|z^*=\tilde{z}})-\left( \sup_{x_1,z_1}|\Delta_n(x_1,z_1)|+2C\right)+O_{\P}\left(n^{-\tau}\right),
    \end{equation*}
    where \( C \) is defined in Assumption \ref{assump:indepnoise}.
\end{corollary}
\noindent
Corollary \ref{cor 1} shows that in the worst-case scenario, the gap between the expected revenue and the maximum expected social welfare is bounded above by \(\sup_{x_1,z_1}|\Delta_n(x_1,z_1)| + 2C\), where \(\sup_{x_1,z_1}|\Delta_n(x_1,z_1)|\) represents the prediction error, and \(2C\) accounts for the value approximation error. Thus, a more accurate estimator \(\hat{\mu}_n\), with a smaller prediction error, results in a higher lower bound for the expected revenue of COAD.

When bidders' values are bounded within $[1,h]$, 
\cite{babaioff2017posting} introduced a deterministic posting price mechanism that guarantees a revenue at least $2\zeta log_{\zeta}h$ fraction of the maximum expected social welfare for any $1 < \zeta \leq h$ when $m^* \geq log_{\zeta}h$.  
In a more general setting where bidders' values are bounded in $[0,h]$, the following corollary demonstrates that COAD  can achieve a similar revenue approximation when the prediction error and value approximation error are well-controlled.

\begin{corollary}\label{theroem 6}
Suppose the bidders' values are bounded within $[0,h]$. For any given item with feature $ z^* = \tilde{z} \in \mathcal{Z} $, and $ 1 < \zeta \leq h $, under  Assumptions \ref{assump:indepnoise} and \ref{assump:consistency}, and assuming that  
\begin{equation*}  
\sup_{x_1,z_1}|\Delta_n(x_1,z_1)|+2C <  \left( 1 - \frac{1}{2\zeta \log_{\zeta} h} \right) h,  
\end{equation*}  
there exists an integer $ M \in \mathbb{N}^+ $ such that for all $ m^* > M $, the expected revenue of COAD satisfies  
\begin{equation*}  
    R_{m^*}^{\text{COAD}|\mathcal{D}}(F_{{v^*}, {x^*}|z^*=\tilde{z}}) > \frac{1}{2\zeta \log_{\zeta} h} W_{m^*}(F_{{v^*}|z^*=\tilde{z}}) + O_{\P}\left(n^{-\tau}\right).  
\end{equation*}  
\end{corollary}
\noindent
Corollary \ref{cor 1} suggests that when $ \sup_{x_1,z_1}|\Delta_n(x_1,z_1)|+2C$ is well-controlled,  COAD can approximate the maximum expected social welfare arbitrarily closely.
We also compare COAD with several traditional auction designs, including Myerson's optimal auction \citep{myerson1981optimal}, the first-price auction \citep{harrison1989theory}, the second-price auction \citep{vickrey1961counterspeculation}, and the second-price auction with a single reserve price \citep{riley1981optimal}. Additional details on these comparisons are provided in the supplementary material.



\section{Real Data Analysis}\label{sec:real_data}
We apply the COAD mechanism to the eBay dataset in Section \ref{sec:datadescrip} and address questions (Q1)-(Q2).
On eBay, a bidder increases their bid if they are not currently the highest bidder and the current price remains below their true valuation. 
The platform effectively implements a second-price auction mechanism \citep{cesa2014regret}, which incentivizes bidders to truthfully report their values. Bidders can use eBay’s automatic bidding system by specifying the maximum amount they are willing to pay—this amount is treated as their true value for the item. 
As a result, for all non-winning bidders, their highest bid typically reflects their true valuation.
To accurately represent bidder valuations, we use each bidder’s highest bid while excluding the winning bid. This is because the winning bid may not reflect the winner's true value, as any bid above a certain threshold yields the same outcome. A similar exclusion strategy has been employed in field experiments to set reserve prices in online advertising auctions \citep{ostrovsky2023reserve}.
In our analysis, we treat seller identity as an item feature, focusing on three main sellers: $\tilde{z}_1$ for `syschannel', $\tilde{z}_2$ for `michael-33', and $\tilde{z}_3$ for `saveking', who account for the majority of auctions. This yields $813$ historical entries, with item features categorized as $\mathcal{Z}=\{\tilde{z}_1, \tilde{z}_2, \tilde{z}_3\}$.
Bidder features include bid time, bidder rating, and average historical bids.  
Each auction involves approximately 12 bidders on average.

To evaluate the revenue guarantees of COAD, we compare its expected revenue against two baselines: (i) the second-price auction, which is widely used in practice—including on eBay—and is known for its incentive compatibility \citep{cesa2014regret}; and (ii) Myerson's auction based on the empirical distribution, following the method of \cite{cole2014sample}, which provides a practical implementation of Myerson’s optimal auction \citep{myerson1981optimal} using historical value data.
In our setting, it is difficult to estimate bidder-specific value distributions for $v^* | (x^*,z^*)$ because individual bidders participate in only a few auctions and bidder features $x^*$ vary continuously. As a result, Myerson’s auction is applied to the conditional distribution of $v^*|z^*$, effectively treating all bidders as symmetric. To implement the empirical Myerson auction, we estimate $F_{v^*|z^*}$ using all observed bidder values for each item.

\begin{figure}[t]
    \centering
    \begin{subfigure}{1\textwidth}
    \centering
    \begin{subfigure}{0.25\textwidth}
        \includegraphics[width=\linewidth]{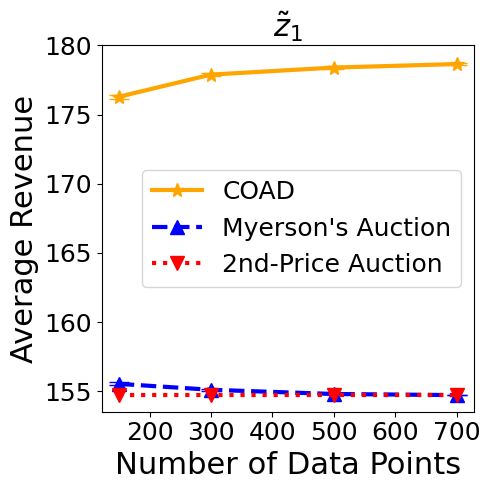}
    \end{subfigure}
    \begin{subfigure}{0.25\textwidth}
        \includegraphics[width=\linewidth]{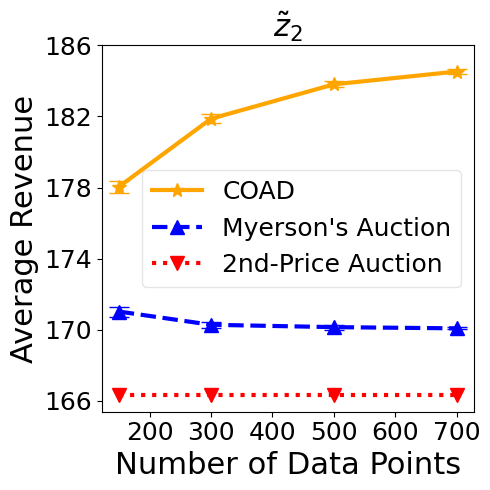}
    \end{subfigure}
    \begin{subfigure}{0.255\textwidth}
        \includegraphics[width=\linewidth]{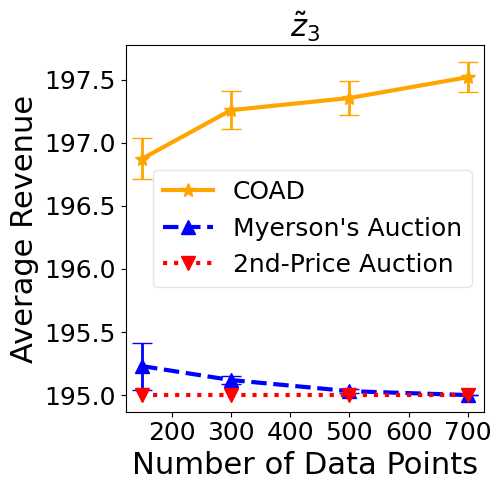}
    \end{subfigure}
    \caption{Average revenue of different auction mechanisms across different numbers of data points.}
    \label{fig:2.1}
    \end{subfigure}

    \vspace{0.3cm} 

    \begin{subfigure}{1\textwidth}
     \centering
       \begin{subfigure}{0.25\textwidth}
        \includegraphics[width=\linewidth]{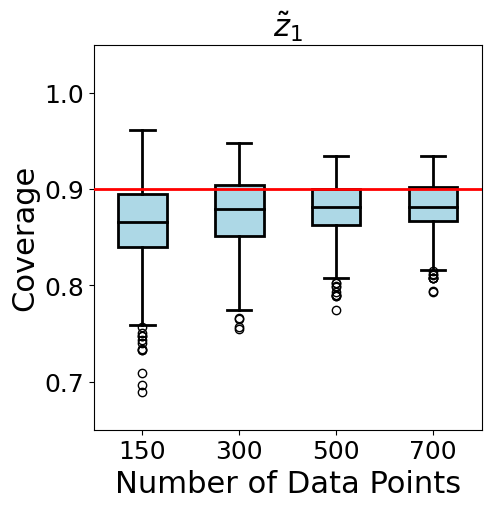}
    \end{subfigure}
    \begin{subfigure}{0.25\textwidth}
        \includegraphics[width=\linewidth]{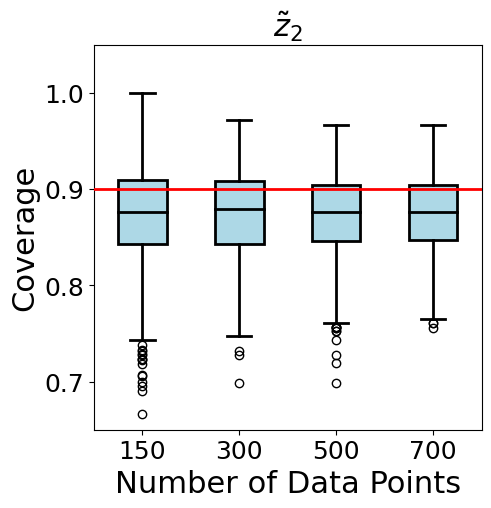}
    \end{subfigure}
    \begin{subfigure}{0.25\textwidth}
        \includegraphics[width=\linewidth]{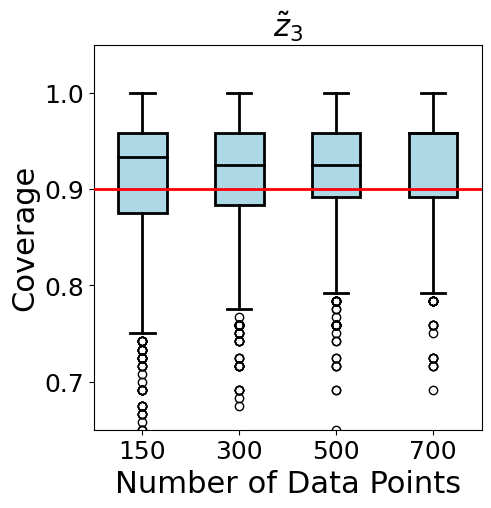}
    \end{subfigure} 
    \caption{The coverage probability of the conformal prediction interval for the true value.}
    \label{fig:2.2}
    \end{subfigure}

    \caption{ \small{Results from Section \ref{sec:real_data}, based on 1000 experiments. (a) Average revenue of different mechanisms for auctions of items with various features \( z^* \), evaluated using different numbers of data points \( N \in \{150, 300, 500, 700\} \). Error bars represent the 95\% confidence interval for the mean. (b) Boxplots of the coverage probability of the conformal prediction interval for the true value, conditioned on different item features \( z^* \), using different numbers of data points \( N \in \{150, 300, 500, 700\} \). The red line denotes the target coverage level of \( 1 - \alpha = 0.9 \).}}
    \label{fig:real:data}
\end{figure}

In our analysis, for each fixed item feature, we extract the corresponding auctions from the dataset and treat them as new auctions for prediction. For each new auction, we randomly sample \( N \) observations from the remaining data to serve as historical data.
We include bidders who have previously placed bids in the new auction. Using conformal prediction conditioned on item features and implemented via quadratic polynomial regression, we apply COAD with a miscoverage level of $\alpha=0.1$.
We also assess the coverage probability of the conformal prediction interval across all item types. We consider various dataset sizes $N\in\{150,300,500,700\}$ and repeat the process $1000$ times for each $N$.

Figure \ref{fig:2.1} shows the average revenue of each mechanism across different items and dataset sizes. In all settings, COAD consistently achieves higher average revenue than both the second-price auction and the empirical Myerson auction. Moreover, as the dataset size increases, COAD’s average revenue steadily improves. This pattern is expected, as larger training sets reduce prediction error and increase revenue, which aligns with our theoretical results in Section \ref{sec:sensitivity}.
Note that the average revenue of the empirical Myerson auction decreases with larger datasets. This is because the mechanism computes a single reserve price based on the estimated value distribution. 
The estimated value distribution may deviate from the true distribution in new auctions, as the i.i.d. assumption often fails to hold in real-world settings. Consequently, there is no guarantee that revenue will improve as the number of data points increases. Moreover, when historical data is limited, the estimated virtual value function—and hence the reserve price—can overfit the empirical distribution, particularly in the upper tail. This overfitting may result in higher revenue when the reserve exceeds the second-highest bid in certain auctions. However, as the dataset grows, the estimation becomes more stable, and the reserve price tends to decrease and better approximate the distribution. This more conservative reserve may fall below the second-highest bid in some auctions, thereby reducing average revenue.

Figure \ref{fig:2.2} presents boxplots of the coverage probability for the conformal prediction intervals across different item features and data sizes. The coverage remains consistently high and becomes more concentrated around the target level as the number of data points increases. These results demonstrate that the conformal prediction method in Section \ref{sec:conformal} provides reliable coverage for the true value, even when Assumptions \ref{assump:iiddata} and \ref{assump:indepbidders} are not fully satisfied, which highlights the robustness of our approach.

Based on both the mechanism design and the numerical results, COAD effectively addresses research questions (Q1) and (Q2).
First, as shown in Figure \ref{fig:2.1}, COAD provides revenue guarantees without requiring large datasets. This is due to its ability to efficiently leverage available data through uncertainty quantification, enabling the construction of bidder-specific reserve prices. Unlike methods that depend on extensive value samples from each bidder to estimate their distribution \citep[e.g.,][]{cole2014sample, roughgarden2016ironing}, COAD makes better use of limited data. In particular, it extracts one data point per bidder per auction, whereas methods like \cite{mohri2016learning} rely only on the highest and second-highest bids, limiting the information available for learning.
Second, the eBay dataset involves heterogeneous bidders and items. COAD effectively handles this variability by incorporating bidder and item features into its design, making it well-suited for real-world platforms where rich feature data is typically available.

\section{Application-Based Simulation Studies}\label{sec:four}
In this section, we provide application-based simulations to evaluate our method by comparing the expected revenue of COAD with that of a second-price auction and Myerson's auction based on the empirical distribution \citep{cole2014sample}. We set the miscoverage level to $\alpha = 0.1$ and randomly generate $N$ iid data points based on the model in \eqref{model_1}, splitting the data equally between calibration and training.  The error bars in the plots indicate the $95\%$ confidence interval for the mean. 

\subsection{A Study Using Neural Networks}
\label{sec:highdsimnn}

The first simulation considers an auction setting where both bidder and item features are $20$-dimensional, i.e.,  $x\in\R^{20}, z\in\mathcal{Z}\subset\R^{20}$. For any $(x,z,v)\  {\sim}\  P$, the item feature $z$ is uniformly selected from $\mathcal{Z}=\{\tilde{z}_1,\dots,\tilde{z}_{q}\}$, where $q=30$ and each $\tilde{z}_{i}$ is independently drawn from $\mathcal{N}(\vec{\boldsymbol{0}}, \boldsymbol{I}_{20})$. The bidder feature $x$ is generated from $\mathcal{N}(\boldsymbol{\mu_x}, \boldsymbol{I}_{20})$, where $\boldsymbol{\mu_x}=(||z||_2^2/{20}, ||z||_2^2/{20},\dots,||z||_2^2/{20})$.
The regression model is,
\begin{equation*}
    \mu(x,z) = e^{\beta_1^\top  x}\cdot (\beta_2^\top  z),
\end{equation*}
with $\beta_1\in\mathbb{R}^{20}, \beta_2\in\mathbb{R}^{20}$, where each entry of $\beta_1$ and $\beta_2$ is independently drawn from Unif$[-0.5,0.5]$. This choice of $\mu(x,z)$ is motivated by prior empirical observations that bidders' values for a given item often follow a log-normal distribution \citep{ostrovsky2023reserve, lahaie2007revenue}. 
The deviation from the expected value is modeled as \( v - \mu(x, z) = e^{\cos^2(x^\top z)} \eta \), which captures the dependency between the value approximation error $v - \mu(x, z)$ and the features $(x,z)$. Here, \( \eta \) is independently drawn from \(\text{Unif}[-1,1]\). 
This setup simulates an online advertising scenario in which ad slots are associated with 30 different keyword types. The item features represent information related to these keywords, while the bidder features capture characteristics of the advertisements, such as the advertiser’s rating and product details.
To fit the model, we use a neural network with two hidden layers containing 128 and 64 neurons, respectively. The network uses LeakyReLU activation functions \citep{maas2013rectifier}, incorporates $L_2$ regularization, applies $30\%$ dropout, and is optimized using the Adam optimizer. The model is trained for $15$ epochs with a batch size of $32$.

\begin{figure}[t!]
    \centering
    \begin{subfigure}[b]{0.235\textwidth}
        \centering
        \includegraphics[width=\textwidth]{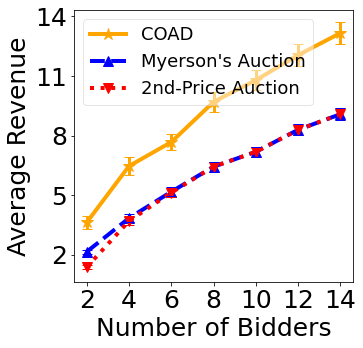}
        \caption{}
        \label{fig:nn-a}
    \end{subfigure}
    \begin{subfigure}[b]{0.24\textwidth}
        \centering
        \includegraphics[width=\textwidth]{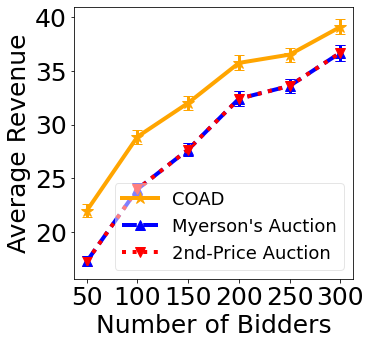}
        \caption{}
        \label{fig:nn-b}
    \end{subfigure}
    \begin{subfigure}[b]{0.25\textwidth}
        \centering
        \includegraphics[width=\textwidth]{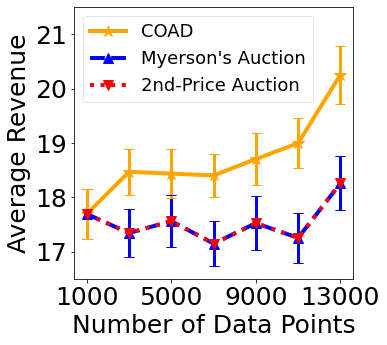}
        \caption{}
        \label{fig:nn-c}
    \end{subfigure}
    \begin{subfigure}[b]{0.247\textwidth}
        \centering
        \includegraphics[width=\textwidth]{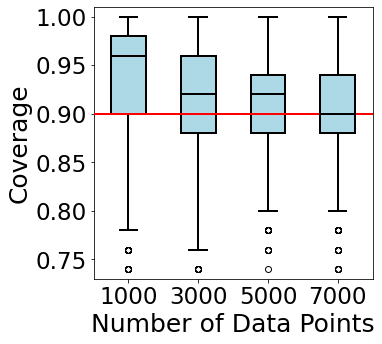}
        \caption{}
        \label{fig:nn-d}
    \end{subfigure}
    \caption{\small{Results from Section \ref{sec:highdsimnn}, based on 1000 experiments for a randomly selected item. (a) Average revenue of different mechanisms, with varying numbers of $m^*\in \{2,4,6,8,10,12,14\}$ and $N=50000$. (b) Average revenue of different mechanisms, with varying numbers of $m^*\in \{50,100,150,200,250,300\}$ and $N=50000$. (c) Average revenue of different mechanisms, with varying numbers of $N\in\{1000, 3000, 5000, 7000, 9000, 11000, 13000\}$ and $m^*=50$. (d) Boxplots of the coverage probability for the true value, with $N\in\{1000,3000,5000,7000\}$ and $m^*=50$. The red line denotes the target coverage level of \( 1 - \alpha = 0.9 \).} }
    \label{fig:combined:high2}
\end{figure}

Figure \ref{fig:nn-a} and \ref{fig:nn-b} illustrate the average revenue of three auction mechanisms under varying numbers of bidders. Figure \ref{fig:nn-a} focuses on scenarios with a small number of bidders, where \( m^* \in \{2,4,6,8,10,12,14\} \), while Figure \ref{fig:nn-b} considers cases with a larger number of bidders, with \( m^* \in \{50,100,150,200,250,300\} \). As \( m^* \) increases, the expected revenue of COAD increases, consistent with Theorem \ref{pp2}. Notably, Myerson’s auction with a single reserve price has a substantial effect only when the number of bidders is small, a trend also observed in \cite{ostrovsky2023reserve}. 
Figure \ref{fig:nn-c} shows the average revenue of the three mechanisms across different dataset sizes when \( m^* = 50 \). COAD consistently outperforms the other two mechanisms in terms of average revenue, with its performance improving as more data become available, leading to lower prediction error, which is consistent with the theoretical results in Section \ref{sec:sensitivity}. 
Figure \ref{fig:nn-d} presents boxplots of the conditional coverage of the prediction intervals across different dataset sizes, showing the empirical distribution of \( \P(v_i^* \in \hat{\mathcal{C}}_{\text{dual}}(x_i^*,z^*) \mid \mathcal{D}, z^*=\tilde{z}) \). As the dataset size increases, the coverage probability becomes more concentrated around the target level  \( 1 - \alpha \), confirming that the uncertainty quantification method in Section \ref{sec:propconformal} achieves the desired conditional coverage.


\subsection{A Study Using Polynomial Regression}\label{poly_high}
The second simulation studies an auction with a higher-dimensional regression model, where both bidder and item features are $100$-dimensional. For any $(x,z,v)\  {\sim}\  P$, the item feature $z$ is uniformly selected from $\mathcal{Z}=\{\tilde{z}_1,\dots,\tilde{z}_{q}\}$ with $q=30$, where each $\tilde{z}_{i}$ is independently drawn from $\mathcal{N}(\vec{\boldsymbol{0}}, \boldsymbol{I}_{100})$. The bidder feature $x$ is generated from $\mathcal{N}(\boldsymbol{\mu_x}, \boldsymbol{I}_{100})$, where $\boldsymbol{\mu_x}=(\|z\|_2^2/{100}, \|z\|_2^2/{100},\ldots,\|z\|_2^2/{100})$. 
The regression model is,
\begin{equation*}
    \mu(x,z) =  (\beta_1^\top x)^2\cdot [\text{sin}^2(\beta_2^\top  z)], 
\end{equation*}
with $\beta_1\in\mathbb{R}^{20}, \beta_2\in\mathbb{R}^{20}$,  where each element of $\beta_1$ and $\beta_2$ is independently drawn from Unif$[-1,1]$. 
The value approximation error is modeled as, $v-\mu(x,z)=e^{\text{cos}^2(x^\top z)}\eta$, where \( \eta \) is independently drawn from a truncated standard normal distribution on $[-1,1]$.
This setup simulates a more complex eBay auction scenario with $30$ distinct item types. Each bidder is characterized by a $100$-dimensional continuous feature vector, and a correlation is introduced between item and bidder features to reflect the realistic scenario in which different types of items appeal to different groups of bidders. We fit the model using quadratic polynomial regression.

\begin{figure}[t!]
    \centering
    \begin{subfigure}[b]{0.25\textwidth} 
        \centering
        \includegraphics[width=\textwidth]{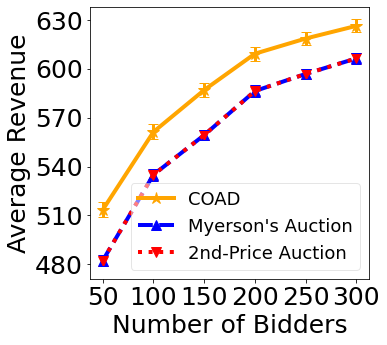}
        \caption{}
        \label{fig:4a}
    \end{subfigure}
    \begin{subfigure}[b]{0.26\textwidth}
        \centering
        \includegraphics[width=\textwidth]{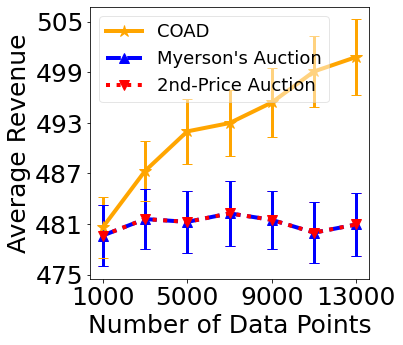}
        \caption{}
        \label{fig:4b}
    \end{subfigure}
    \begin{subfigure}[b]{0.25\textwidth}
        \centering
        \includegraphics[width=\textwidth]{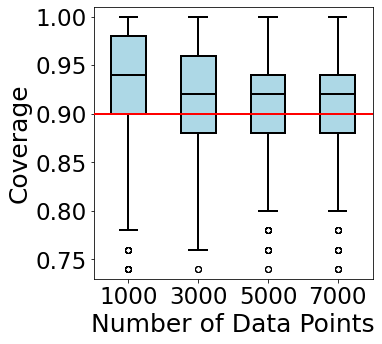}
        \caption{}
        \label{fig:4c}
    \end{subfigure}
    \caption{\small{Results from Section \ref{poly_high}, based on 1000 experiments for a randomly selected item. (a) Average revenue of different mechanisms, with varying numbers of $m^*\in \{50,100,150,200,250,300\}$ and $N=20000$. (b) Average revenue of different mechanisms, with varying numbers of $N\in\{1000, 3000, 5000, 7000, 9000, 11000, 13000\}$ and $m^*=50$. (c) Boxplots of the coverage probability for the true value, with varying numbers of $N\in\{1000,3000,5000,7000\}$ and $m^*=50$. The red line denotes the target coverage level \( 1 - \alpha = 0.9 \).}}
    \label{fig:combined}
\end{figure}

Figure \ref{fig:4a} shows the average revenue of three auction mechanisms as the number of bidders $m^*$ varies. As $m^*$ increases, the expected revenue of COAD also increases, consistent with Theorem \ref{pp2}. 
Figure \ref{fig:4b} shows the average revenue of the three mechanisms for different dataset sizes with $m^*=50$. In scenarios with a large number of bidders, the empirical Myerson auction with a single reserve price performs similarly to the second-price auction. However, COAD consistently achieves a higher average revenue than both mechanisms.
Figure \ref{fig:4c} presents boxplots of the conditional coverage probabilities of the conformal prediction intervals across different dataset sizes.  The results confirm that the method introduced in Section \ref{sec:propconformal} achieves conditional coverage at the level \( 1 - \alpha \).

\section{Related Works}
\label{sec:relatedwork}
We briefly review related work from several fields, including optimal auction design, posted-price auction, mechanism design, and statistical uncertainty quantification.

\vspace{0.1in}
\noindent
\textbf{Optimal auction design.}
Most theoretical studies on revenue-maximizing auctions trace back to the foundational works of \cite{myerson1981optimal} and \cite{riley1981optimal}. In particular, \cite{myerson1981optimal} integrated the Vickrey-Clarke-Groves (VCG) mechanism with a reserve price under the assumption that bidders' values are drawn from a known regular distribution.
Existing literature explores two main approaches to revenue-maximizing auctions when value distributions are unknown. One approach estimates empirical distributions using historical data \citep{cole2014sample, huang2015making, morgenstern2015pseudo, devanur2016sample, roughgarden2016ironing, guo2019settling}. Another approach leverages statistical properties of bidders' value distributions, such as the mean, variance, or quantiles, to design revenue-maximizing auctions \citep{azar2013parametric, azar2013optimal}.
Our COAD mechanism differs from these methods by incorporating bidder and item features and using a distribution-free, prediction interval-based method for revenue maximization.

\vspace{0.1in}
\noindent
\textbf{Posted-price auction mechanism.} 
We have designed a bidder-specific reserve price mechanism, which relates to posted-price auction mechanisms commonly used in online auctions \citep{blum2004online}. In these auctions, auctioneers set prices, and buyers decide whether to accept or reject them.
The reserve price determined by COAD can be used to set prices in a sequential posted-price mechanism. However, unlike traditional approaches, COAD requires bidders to submit their bids before the price is revealed.
Previous research on posted-price mechanisms with \emph{known} priors \citep[e.g.,][]{chawla2009sequential} has focused on setting prices based on prior knowledge of buyers' value distributions.
For cases where the distribution is \emph{unknown}, \cite{balcan2008reducing}  introduced a strategy that leverages buyers' features. Our mechanism extends this approach by incorporating both bidder and item features. Additionally, \cite{babaioff2017posting} developed a method that guarantees a constant fraction of the optimal revenue under the assumption of a known bounded support. In contrast, our method provides revenue guarantees without requiring prior knowledge of the distribution’s support, making it more adaptable to real-world auction settings.

\vspace{0.1in}
\noindent
{\bf{Learning and robustness in mechanism design.}} 
The challenge of private and unknown preferences has driven extensive research on learning-based mechanism designs, focusing on auctions that are robust to errors in bidder value distributions without relying on Bayesian assumptions \citep{cai2017learning, brustle2020multi}. There is also growing interest in integrating machine learning into market designs \citep{dai2021learning, dai2021learningjmlr, dai2022incentive}. In this paper, we use high-confidence prediction intervals, developed through conformal prediction on historical data, ensuring our auction mechanism is both practical and capable of achieving high expected revenue.

\vspace{0.1in}
\noindent
\textbf{Conformal inference.} 
The conformal prediction framework, introduced by \citet{vovk2005algorithmic}, enables effective uncertainty quantification through prediction intervals with marginal coverage guarantees in finite samples, without assumptions about the data-generating process \citep{lei2013distribution, lei2018distribution}. While achieving conditional coverage alongside marginal coverage is challenging \citep{vovk2012conditional, foygel2021limits}, \cite{gibbs2023conformal} recently proposed a method for group-conditional coverage and coverage under covariate shift.  
In our study, we extend the application of conformal prediction to auction design in economics, presenting a novel use of this method.

\section{Conclusion}\label{sec:six}
We introduce Conformal Online Auction Design (COAD), a novel data-driven approach to online auction design that maximizes revenue while ensuring incentive compatibility.  COAD leverages historical data to predict each bidder’s valuation with uncertainty quantification and applies bidder-specific reserve prices, addressing the dynamic and heterogeneous nature of online auctions. This flexibility makes COAD particularly well-suited for applications such as eBay auctions and online advertising, where bidder features and values can vary widely. By employing conformal prediction techniques, COAD operates effectively with any finite sample of historical data without requiring knowledge of the underlying value distributions. Our real-world applications and application-based simulations demonstrate that COAD outperforms traditional revenue benchmarks, providing a compelling alternative to existing online auction designs.

COAD can be extended to various settings. One natural extension is to online auctions where bidders participate repeatedly under budget constraints. In settings like eBay auctions in our real data analysis, bidders typically engage in only a small number of auctions to purchase a specific item. In such cases, where a bidder aims to win the auction only once and makes a single payment, their budget represents their true valuation. This allows COAD to directly estimate their actual values without additional budget considerations. However, in online advertising platforms such as Meta and Google, each user visit generates an ad opportunity for advertisers to bid on. Given the continuous stream of ad placements, advertisers rely on auto-bidding agents to optimize their bids while managing their budgets \citep{wu2018budget}. In these environments, COAD can be adapted by constructing prediction intervals for strategic bids. Since bid distributions vary with budget constraints, COAD can incorporate real-time budget information as a feature in the prediction model. The lower confidence bounds of these prediction intervals can then be used to set bidder-specific reserve prices. If budget constraints are not directly observable to auctioneers, they can be inferred using other bidder features and incorporated into COAD to improve bid estimates.

Another extension of COAD involves relaxing its underlying assumptions.  Assumptions \ref{assump:iiddata} and \ref{assump:indepbidders} are introduced to ensure that conformal prediction methods provide exact coverage guarantees for prediction intervals, which in turn ensures theoretical revenue guarantees. However, in practice, these assumptions are not always necessary. When the prediction error of the machine learning model and the value approximation error are small, COAD can still achieve high revenue guarantees without relying on these assumptions. This is supported by Corollaries \ref{cor 1} and \ref{theroem 6}, where we establish COAD’s revenue guarantee in the worst-case scenario without requiring Assumptions \ref{assump:iiddata} and \ref{assump:indepbidders}. Additionally, in cases where the data-generating distribution evolves over time with general dependence, an alternative method for constructing prediction intervals is adaptive conformal inference \citep[e.g.,][]{gibbs2021adaptive}. This approach iteratively adjusts the interval length based on the discrepancy between the observed coverage and the target level $1-\alpha$. Specifically, if the achieved coverage exceeds the target, the interval length is reduced, whereas if it falls short, the interval is expanded. This method is particularly useful when large volumes of auction data are available for each item.



\section*{Acknowledgments}
We thank Professor Mike Jordan for suggesting the importance of uncertainty quantification in mechanism design and for his valuable feedback.
This work was partially supported by NIH grants R01DK142026 and P2C-HD041022, a Merck Research Award, and a Hellman Fellowship Award.


\baselineskip=17pt
\bibliography{conformal}

\newpage
\appendix
\normalsize
\section*{Appendix}
\noindent 
This appendix contains additional details of our method, the proofs, and additional simulations.   Section
\ref{A.detail} provides a detailed explanation of the construction of the conformal prediction interval discussed in Section \ref{sec:conformal}, 
Section \ref{sec:compare} compares COAD with other traditional auction designs.
Section \ref{app:proofs} contains all the proofs of the theories presented in the paper, 
and Section \ref{app:add_exp} presents additional experimental results.

\section{Details of Constructing Conformal Prediction Intervals}\label{A.detail}
The conformal prediction method in this paper is based on \cite{gibbs2023conformal} and provides a conditional guarantee for the valuation of each bidder in $[m^*]$. The construction procedure consists of two main steps. Step 1 is to generate the primal prediction interval $\hat{\mathcal{C}}_{\text{primal}}$. Step 2 is to redefine $\hat{\mathcal{C}}_{\text{primal}}$ in terms of the dual formulation and obtain the dual prediction band $\hat{\mathcal{C}}_{\text{dual}}$.

Given the conformity score function $\mathcal{S}:\mathcal{X}\times \mathcal{Z}\times \mathbb{R}_{\geq 0}\to \mathbb{R}$ defined as $\mathcal{S}(\{x,z\},v)=|v-\hat{\mu}_n(x,z)|$ for any $(x,z,v)\in\mathcal{X}\times\mathcal{Z}\times \mathbb{R}_{\geq 0}$, let $\tilde{S}_{i}$ denote the score $\mathcal{S}(\{x_i^*,z^*\},v_i^*)$ for $i\in[m^*]$, and $S_1,\dots,S_{n}$ denote the calibration scores $\mathcal{S}(\{x_1,z_1\},v_1),\dots, \mathcal{S}(\{x_n,z_n\},v_n)$ of all the calibration data. For each $i\in[m^*]$, from Assumption \ref{assump:indepbidders}, $(x_i^*,z^*,v_i^*)$ is iid of the $D_{\text{cal}}$, so that $\tilde{S}_{i}, S_1,\dots,S_n$ are also iid random variables.

Let $\mathcal{G}=\Big\{\{\mathcal{X},\tilde{z}_1\},\dots,\{\mathcal{X},\tilde{z}_{q}\}\Big\}$, which is a finite collection of groups in $2^{\mathcal{X}\times\mathcal{Z}}$, and it divides the domain $\mathcal{X}\times\mathcal{Z}$ into $q$ pieces. And let 
\begin{equation*}
\begin{split}
      \mathcal{F}&=\Big\{(x,z)\longmapsto\sum_{G\in\mathcal{G}}\beta_G\mathbb{I}\Big\{\{x,z\}\in G\Big\} \ \Big|\  \beta_G\in \R,\  \forall G\in \mathcal{G} \Big\},
\end{split}
\end{equation*}
which is a linear function space spanned by the identification function over $\mathcal{G}$.  

The augmented quantile regression estimate for the bidder $i\in[m^*]$ and the item 
with feature $z^*$ is defined as:
\begin{equation}\label{gs}
    \hat{g}^i_{S}=\argmin_{g\in\mathcal{F}}\frac{1}{n+1}\sum^n_{j=1}\ell_{\alpha}(g(x_j,z_j),S_j)+\frac{1}{n+1}\ell_{\alpha}(g(x_i^*,z^*),S).
\end{equation}
where
$\ell_{\alpha}$ is the "pinball" loss and $\alpha\in(0,1)$:
\begin{equation*}
 \ell_{\alpha}(\theta,R)= \begin{cases}
(1-\alpha)(R-\theta)\quad \text{if } R\geq \theta, \\
\alpha(\theta-R)\quad \text{if } R< \theta.
\end{cases} 
\end{equation*}
Then, take the prediction interval for $v_i^*$, which is the value of the bidder $i$, to be 
\begin{equation}\label{ps}
    \hat{\mathcal{C}}_{\text{primal}}(x_i^*,z^*)=\{v: \mathcal{S}(\{x_i^*,z^*\},v)\leq \hat{g}^i_{\mathcal{S}(\{x_i^*,z^*\},v)}(x_i^*,z^*)\}.
\end{equation}
From \emph{group-conditional coverage} guarantee in \cite{gibbs2023conformal}, we have that,
\begin{equation}\label{c_prime}
    \mathbb{P}(v_i^*\in \hat{\mathcal{C}}_{\text{primal}}(x_i^*,z^*)\ |\ z^*=\tilde{z})\geq 1-\alpha \quad \text{for all } \tilde{z}\in \mathcal{Z}.
\end{equation}

Before getting the dual formulation, first, we note from equation \eqref{gs} that $\hat{g}^i_S$ is the optimal solution of the following unconstrained optimization problem when $x=x^*_i, z=z^*$: 
\begin{equation}\label{unc}
    \mathop{\text{minimize}}\limits_{g\in\mathcal{F}}\frac{1}{n+1}\sum^n_{j=1}\ell_{\alpha}(g(x_j,z_j),S_j)+\frac{1}{n+1}\ell_{\alpha}(g(x,z),S).
\end{equation}
Let $p_j=(S_j-g(x_j,z_j))\mathbb{I}(S_j\geq g(x_j,z_j))$, and $q_j=(g(x_j,z_j)-S_j)\mathbb{I}(S_j< g(x_j,z_j))$, for $j=1,2,\dots,n$. Let $p_{n+1}=(S-g(x,z))\mathbb{I}(S\geq g(x,z))$, and $q_{n+1}=(g(x,z)-S)\mathbb{I}(S< g(x,z))$. Denote $\boldsymbol{p}=(p_1,p_2,\dots,p_{n+1})$, and $\boldsymbol{q}=(q_1,q_2,\dots,q_{n+1})$. Then the problem \eqref{unc} can be re-formulated into  the following (relaxed) constrained  optimization problem: 
\begin{equation}\label{constrain}
\begin{split}
\mathop{\text{minimize}}\limits_{\boldsymbol{p},\boldsymbol{q}\in\mathbb{R}^{n+1},\ g\in\mathcal{F} }&\quad\sum^{n+1}_{j=1}\big[(1-\alpha)p_j+\alpha q_j\big],\\
\text{sbject to } &\quad S_j-g(x_j,z_j)-p_j+q_j=0,\ j=1,2,\dots, n,\\
&\quad S-g(x,z)-p_{n+1}+q_{n+1}=0,\\
&\quad p_j,q_j\geq 0,\ j=1,2,\dots, n+1.
\end{split}
\end{equation}
For the constrained problem \eqref{constrain}, we can use the Lagrange multipliers $\boldsymbol{\eta}=(\eta_1,\eta_2,\dots,\eta_{n+1})$, $\boldsymbol{\gamma}=(\gamma_1,\gamma_2,\dots,\gamma_{n+1})$, $\boldsymbol{\xi}=(\xi_1,\xi_2,\dots,\xi_{n+1})$ and consider its Lagrangian
\begin{equation*}
\begin{split}
\mathop{\text{minimize}}\limits_{\boldsymbol{p},\boldsymbol{q},\boldsymbol{\eta}, \boldsymbol{\gamma}, \boldsymbol{\xi}\in\mathbb{R}^{n+1},\ g\in\mathcal{F} }
\mathcal{L}(g,\boldsymbol{p},\boldsymbol{q},\boldsymbol{\eta}, \boldsymbol{\gamma}, \boldsymbol{\xi})=&\quad\sum^{n+1}_{j=1}\big[(1-\alpha)p_j+\alpha q_j\big] + \sum_{j=1}^n \eta_j(S_j-g(x_j,z_j)-p_j+q_j) \\
&+ \eta_{n+1}(S-g(x,z)-p_{n+1}+q_{n+1})-\sum_{j=1}^{n+1}(\gamma_jp_j+\xi_jq_j).
\end{split}
\end{equation*}
The Karush-Kuhn-Tucker (KKT) conditions for this Lagrangian also include
\begin{equation*}
\begin{split}
\text{stationary equations  }&
\Delta_{\boldsymbol{p}} \mathcal{L} = \boldsymbol{0},\  \Delta_{\boldsymbol{q}} \mathcal{L} = \boldsymbol{0},\ 
\Delta_g \mathcal{L} = \boldsymbol{0},\\
\text{dual feasibility  }&\gamma_j,\xi_j\geq 0,\ j=1,2,\dots, n+1,\\
\text{complementary slackness  }& \gamma_j p_j=0, \ \xi_jq_j=0,\ j=1,2,\dots, n+1.
\end{split}
\end{equation*}
Now if we focus on finding the optimal solution for $\boldsymbol{\eta}$, from the stationary equations of $\boldsymbol{p}, \boldsymbol{q}$, we have that,
\begin{equation}\label{eq:17}
\begin{split}
    \boldsymbol{\gamma} &= (1-\alpha)\cdot  \boldsymbol{1} -\boldsymbol{\eta},\\
    \boldsymbol{\xi} &= \alpha\cdot \boldsymbol{1}+\boldsymbol{\eta}.
\end{split}    
\end{equation}
By \eqref{eq:17} and dual feasibility, it follows that the constraint on  $\boldsymbol{\eta}$ is defined as, 
\begin{equation*}
    -\alpha\cdot \boldsymbol{1} \leq\boldsymbol{\eta}\leq (1-\alpha)\cdot \boldsymbol{1}.
\end{equation*}
Consequently, we can derive the dual formulation with respect to  $\boldsymbol{\eta}$,
\begin{equation}\label{dual}
\begin{split}
\mathop{\text{maximize}}\limits_{\boldsymbol{\eta}\in\R^{n+1}}&\quad{\sum^n_{j=1}\eta_j S_j+\eta_{n+1}S+\min_{g\in\mathcal{F}}}\{-\sum^{n}_{j=1}\eta_j g(x_j,z_j)- \eta_{n+1}g(x,z)\},\\
\text{sbject to } &\quad -\alpha\leq\eta_j\leq 1-\alpha,\ 1\leq j \leq n+1.
\end{split}
\end{equation}

It is shown in \cite{gibbs2023conformal} that the primal-dual pair \eqref{constrain} and \eqref{dual} satisfies the strong duality by Slater's condition; thus, the optimal primal-dual solution will also satisfy the KKT conditions. If we let $(\hat{g}^i_S, \boldsymbol{\eta}_i^S)$ denote the primal-dual solution for $(g,\boldsymbol{\eta})$ given $S$ when $x=x_i^*,z=z^*$, from \eqref{eq:17} and the complementary slackness conditions when $j=n+1$, we obtain
\begin{equation}\label{eq:19}
\eta^S_{i,n+1}= \begin{cases}
-\alpha,\quad\quad\quad\ \ \ \  \text{if }S< \hat{g}_S^i(x_i^*,z^*), \\
[-\alpha,1-\alpha],\quad \text{if }S= \hat{g}_S^i(x_i^*,z^*), \\
1-\alpha,\ \ \ \ \ \ \ \ \ \ \  \text{if }S> \hat{g}_S^i(x_i^*,z^*).
\end{cases} 
\end{equation}

From \eqref{eq:19} we know that checking whether $S\leq \hat{g}^i_S(x_i^*,z^*)$ is nearly equivalent to check $\eta^S_{i,n+1}<1-\alpha$. Intuitively, we can go back to \eqref{ps} by letting $S=\mathcal{S}(\{x_i^*,z^*\},v)$ in \eqref{eq:19}, so that we can replace $\hat{\mathcal{C}}_{\text{primal}}(x_i^*,z^*)$ by the dual prediction interval,
\begin{equation*}  
    \hat{\mathcal{C}}_{\text{dual}}(x_i^*,z^*)=\{v: \eta^{\mathcal{S}(\{x_i^*,z^*\},v)}_{i,n+1}<1-\alpha\}.
\end{equation*}
We can find that $ \hat{\mathcal{C}}_{\text{dual}}(x_i^*,z^*)$ is obtained from $\hat{\mathcal{C}}_{\text{primal}}(x_i^*,z^*)$ by removing a negligible portion of the points $v$ that lie on the boundary, 
\begin{equation*}
    \{v: \mathcal{S}(\{x_i^*,z^*\},v)= \hat{g}^i_{\mathcal{S}(\{x_i^*,z^*\},v)}(x_i^*,z^*)\}.
\end{equation*}
The following theorem states that $S\mapsto\eta^S_{n+1}$ is non-decreasing.
\begin{theorem}\label{cite:theo}
(\cite{gibbs2023conformal} Theorem 4).
    For all maximizes $\{\eta^S_{n+1}\}_{S\in\mathbb{R}}$ of \eqref{dual}, $S\mapsto\eta^S_{n+1}$ in non-decreasing in $S$.
\end{theorem}
Using Theorem \ref{cite:theo}, if we compute a ${S}^*_i$ which is the largest value of $S$ such that $\eta^S_{i,n+1}<1-\alpha$, 
we can rewrite the dual prediction interval as 
\begin{equation*}
    \hat{\mathcal{C}}_{\text{dual}}(x_i^*,z^*)=\{v: \mathcal{S}(\{x_i^*,z^*\},v)\leq {S}^*_i\}=[\hat{\mu}_n(x^*_i,z^*)-{S}^*_i,\hat{\mu}_n(x^*_i,z^*)+{S}^*_i].
\end{equation*}
\change{To compute \( S_i^* \), a binary search over \( S \) can be used to find the largest value such that \( \eta^S_{i, n+1} \) remains below the desired cutoff \( (1-\alpha) \) \citep{gibbs2023conformal}. We provide the binary search procedure in Algorithm \ref{alg:binary_search}.  Additionally, a Python package \textsf{conditionalconformal} is available or constructing conditional conformal prediction intervals.} Similar to the \eqref{c_prime}, it also follows:
\begin{equation*}
      \mathbb{P}(v_i^*\in \hat{\mathcal{C}}_{\text{dual}}(x_i^*,z^*)\ |\ z^*=\tilde{z})\geq 1-\alpha \quad \text{for all } \tilde{z}\in \mathcal{Z}.
\end{equation*}

\begin{algorithm}[t!]
\caption{ \normalsize{{Binary Search Computation of $S_i^*$}}}
\begin{algorithmic}[1]
\State  \normalsize{\textbf{Input:} Observed data $\{(x_1,z_1, S_1),\dots,(x_n,z_n,S_n)\} \cup \{x^*_i,z^*\}$, numerical error tolerance $\epsilon$, range $[a,b]$ for $S$ ($a, b$ = \texttt{None} indicates that no bounds are known for $S$).}
\State \textbf{if} $b =$ \texttt{None} \textbf{then} $b = \max_{1 \leq i \leq n} S_i$\;
\State \textbf{if} $a =$ \texttt{None} \textbf{then} $a = 0$\;
\State \textbf{while} $b-a \geq \epsilon$ \textbf{do} 
\State \quad \textbf{if} $\eta^{(a+b)/2}_{n+1} < 1-\alpha$ \textbf{then} $a = (a+b)/2$;
\State \quad \textbf{else}  $b = (a+b)/2$.

\State \textbf{Output:} $a$.
\end{algorithmic}
\label{alg:binary_search}
\end{algorithm}
In our setting, based on the definition of $\mathcal{F}$, for any $g\in\mathcal{F}$, the value of $g(x,z)$ is unrelated to $x$, so that $S^*_1=S^*_2=\dots=S^*_{m^*}$, we can denote them simply by $S^*$.


\section{Comparisons with Alternative Auction Designs}\label{sec:compare}
We now compare the COAD mechanism in Algorithm \ref{alg:coad} with 
with several alternative auction designs: Myerson’s optimal auction \citep{myerson1981optimal}, the first-price auction \citep{harrison1989theory}, the second-price auction (or VCG mechanism for single items \citep{vickrey1961counterspeculation}), and the second-price auction with a single item-specific reserve price \citep{riley1981optimal, cesa2014regret, mohri2016learning}.

\vspace{0.1in}
\noindent
\textbf{Comparison with optimal auctions} Myerson’s optimal auction cannot be applied in scenarios where the auctioneer lacks prior knowledge of the bidders' value distributions. In contrast, Theorem \ref{prop1} demonstrates the practical utility of COAD in these settings by guaranteeing revenues.

\vspace{0.1in}
\noindent
\textbf{Comparison with first-price and second-price auctions}
Unlike the first-price auction, which is not incentive-compatible, the second-price auction mechanism is known for its incentive compatibility. The COAD mechanism effectively bridges the first-price and second-price auction mechanisms. By Theorem \ref{prop:incentive}, COAD ensures truthful bidding, similar to the second-price auction, but typically achieves higher revenue. This higher revenue is observed particularly when the highest bidder's reserve price lies between the highest and the second-highest bids.
To illustrate this point, consider the following example.
\begin{example}[Comparison with second-price auctions]
If $m^*=1$, then the revenue from a second-price auction is 0, while the revenue from COAD is that $\max\{0,\hat{v}_1^L\}\geq 0$. If $m^* \geq 2$, without loss of generality, let bidder $i=1$ have the highest value $v_1^*$ and bidder $i=2$ the second highest $v_2^*$. 
Assume that $v_1^* - v_2^* \geq  2$ and that the residuals $\varepsilon$'s are standardized within the range $[-1,1]$. If the prediction rule is sufficiently accurate such that $\hat{\mu}_n = \mu$ and $S^*$ is equal to  $q_{\alpha/2}$, the $(1-\alpha/2)$ quantile of the distribution of $\varepsilon$. Then the probability that the revenue of COAD surpasses that of a second-price auction is given by $\P(v_1^*\geq v_1^L>v_2^*)\geq 1-\alpha/2$.
\end{example}

\vspace{0.1in}
\noindent
\textbf{Comparison with item-specific reserve price} The item-specific reserve price is characterized
by a single reserve price applicable to all bidders for a given item. In contrast, the proposed COAD mechanism employs bidder-specific reserve prices, meaning the reserve price is tailored to each bidder for the same item. We present the following example to illustrate the potential for increased revenue when using bidder-specific reserve prices.
\begin{example}[Comparison with item-specific reserve pricing auctions] \label{example:2}
Consider the distribution $F_{v^*|z^*=\tilde{z}}$ defined as \change{$\mathbb{I}\{ x \geq 1 \} \left( 1 - \frac{1}{H} \right) + \mathbb{I}\{ x \geq H \} \frac{1}{H}$, where $H>1$ is a large constant.}
As shown in \citet{roughgarden2016ironing}, the best item-specific reserve price auction for this distribution achieves an expected revenue of at most $1$. 
\change{However, according to Corollary \ref{cor 1}, the COAD mechanism achieves expected revenue satisfying
\begin{equation*}
\begin{aligned}
    R_{m^*}^{{\text{COAD}}|\mathcal{D}}(F_{{v^*}, {x^*}|z^*= \tilde{z}})
    & \geq H-(H-1)\left(1-\frac{1}{H}\right)^{m^*}-\left( \sup_{x_1,z_1}|\Delta_n(x_1,z_1)|+2C\right).
\end{aligned}
\end{equation*} 
As \( m^* \to \infty \), the term $\left(1-\frac{1}{H}\right)^{m^*} \to 0$, and the lower bound on COAD’s revenue approaches
$H-\left( \sup_{x_1,z_1}|\Delta_n(x_1,z_1)|+2C\right)$. Therefore, if we choose \( H > \sup_{x_1,z_1} |\Delta_n(x_1,z_1)| + 2C + 1 \), the expected revenue of COAD exceeds that of the optimal item-specific reserve price auction.}

\end{example}
Bidder-specific reserve prices have been successfully implemented in real-world auctions to increase revenue. For example, search engines like Google and Yahoo! use advertiser-specific features to set unique reserve prices, ensuring that bids and prices per click have varying minimum thresholds for different advertisers. This strategy not only boosts revenue but also encourages advertisers to place higher-quality ads \citep{even2008position}. This demonstrates the practical relevance of COAD’s bidder-specific reserve price strategy.


\section{Proofs of Main Results}\label{app:proofs}
\noindent 
In this section, we provide technical proof for the main results. 
Section \ref{B.1} provides the proof of Proposition \ref{prop:length}; Section \ref{B.3} includes the proof of Theorem \ref{prop:incentive}; Section \ref{A1.3.1} gives the proof of Theorem \ref{pp2}; Section \ref{A1.2} contains the proof of Theorem \ref{prop1}; Section \ref{proof:cor1} provides the proof of Corollary \ref{cor 1}; Section \ref{proof:cor2} contains the proof of Corollary \ref{theroem 6}.

\subsection{Proof of Proposition \ref{prop:length}}\label{B.1}
\begin{proof}
    From the construction procedure of the dual prediction interval outlined in Section \ref{A.detail},  it is evident that $0<S^*\leq \hat{g}_{S^*}^i(x_i^*,z^*)<\infty$ for any $i\in[m^*]$. By referring to the definition of the augmented quantile regression estimate $\hat{g}^i_{S^*}$, as provided in \eqref{gs}, we can infer $\hat{g}_{S^*}^i(x_i^*,z^*)\leq \max\{S_1, S_2,\dots, S_n, S^*\}$. 
    Given that  $ S^* < \infty $, it can be deduced that at least one of the points among $ z_1, \dots, z_n $ is equal to $ z^* $. This is because otherwise,  $\mathcal{S}(\{x_i^*,z^*\},v)= \hat{g}^i_{\mathcal{S}(\{x_i^*,z^*\},v)}(x_i^*,z^*)$ always holds for any $v\in\R_{\geq 0}$, therefore resulting in $S^*=\infty$. 
    Additionally, considering that $ S^* \leq \hat{g}_{S^*}^i(x_i^*, z^*) $ together with the definition of the pin-loss function, we have that $ S^* \leq \max\{S_1, S_2, \dots, S_n\} $. 
    Hence, we can derive that,
    \change{\begin{equation}\label{eq:27}
    \begin{aligned}
        S^* & \leq\max\{S_1,S_2,\dots, S_n\}\\
        & =\max\{|v_1-\hat{\mu}_n(x_1,z_1)|,|v_2-\hat{\mu}_n(x_2,z_2)|,\dots, |v_n-\hat{\mu}_n(x_n,z_n)|\}\\
        &=\max_{1\leq j \leq n}\{v_j-\hat{\mu}_n(x_j,z_j)\}.\\     
    \end{aligned}
    \end{equation}

    Since it holds that,
    \begin{equation}\label{jasa 26}
    \begin{split}
        \max_{1\leq j \leq n}\{v_j-\hat{\mu}_n(x_j,z_j)\}&=\max_{1\leq j \leq n}\{v_j-\mu(x_j,z_j)+\mu(x_j,z_j)-\hat{\mu}_n(x_j,z_j)\}\\
        \leq&\max_{1\leq j \leq n}\{v_j-\mu(x_j,z_j\}+\max_{1\leq j \leq n}\{\mu(x_j,z_j)-\hat{\mu}_n(x_j,z_j)\}\\
        \leq&\sup_{(x, z, v) \sim P} |v - \mu(x, z)| +\max_{1\leq j \leq n}\{|\mu(x_j,z_j)-\hat{\mu}_n(x_j,z_j)|\}\\
        =&C+\max_{1 \leq j \leq n} |\Delta_n(x_j,z_j)|.
    \end{split}
    \end{equation}

    Combining \eqref{eq:27} with \eqref{jasa 26}, we obtain
    \[S^* \leq \max_{1 \leq j \leq n} |\Delta_n(x_j,z_j)|  + C.\]
    
    This completes the proof of Proposition \ref{prop:length}.
    }
\end{proof}

\subsection{Proof of Theorem \ref{prop:incentive}}\label{B.3}
\begin{proof}
In the proposed COAD mechanism, for any bidder $i\in[m^*]$,
$ a_i(\vec{v}^*,\vec{x}^*,z^*)=1$ implies,
\begin{equation*}
    c_i(v_i^*,x_i^*,z^*)=\max_{k\in[m^*]}c_k(v_k^*,x_k^*,z^*)>0.
\end{equation*}
In other words, a bidder who has a \change{pseudo-virtual value} greater than all other bidders is guaranteed to win the item. Conversely, if a bidder's \change{pseudo-virtual value} is less than that of any other bidder, that bidder will not win the item. That is,
\begin{equation*}
a_i(b_i^*, \vec{v}^*_{-i},\vec{x}^*,z^*)= \begin{cases}
1,\quad b_i^*>  r_i(\vec{v}^*_{-i},\vec{x}^*,z^*), \\
0,\quad b_i^*<r_i(\vec{v}^*_{-i},\vec{x}^*,z^*).
\end{cases}
\end{equation*}
The well-known envelope formula \citep{myerson1981optimal} indicates that if the payment is, 
\begin{equation}\label{envelope}
    p_i(\vec{v}^*,\vec{x}^*,z^*)=a_i(\vec{v}^*,\vec{x}^*,z^*) v_i^*-\int^{v_i^*}_0 a_i(b_i^*, \vec{v}_{-i}^*,\vec{x}^*,z^*)db_i^*,
\end{equation}
for any $i\in[m^*],  \vec{v}^*\in\mathbb{R}^{m^*}_{\geq 0}, (\vec{x}^*,z^*)\in \mathcal{X}^{m^*}\times\mathcal{Z}$, then the mechanism is both IC and IR.  Now we plug in the allocation rule into the envelope formula \eqref{envelope}. Note that 
\begin{equation*}
\int^{v_i^*}_0 a_i(b_i^*, \vec{v}_{-i}^*,\vec{x}^*,z^*)db_i^*= \begin{cases}
v^*_i- r_i(\vec{v}^*_{-i},\vec{x}^*,z^*),\quad v^*_i\geq  r_i(\vec{v}^*_{-i},\vec{x}^*,z^*), \\
0,\quad \quad\quad\quad\quad  \quad\quad\quad\   v^*_i\leq r_i(\vec{v}^*_{-i},\vec{x}^*,z^*).
\end{cases}
\end{equation*}
On the one hand, if the allocation rule $a_i(\vec{v}^*,\vec{x}^*,z^*)=1$,  it follows that $v^*_i\geq  r_i(\vec{v}^*_{-i},\vec{x}^*,z^*)$, so that we can deduce $p_i(\vec{v}^*,\vec{x}^*,z^*)=v^*_{i}-(v^*_i- r_i(\vec{v}^*_{-i},\vec{x}^*,z^*))=r_i(\vec{v}^*_{-i},\vec{x}^*,z^*)$. On the other hand, if $a_i(\vec{v}^*,\vec{x}^*,z^*)=0$, it holds that  $v^*_i\leq  r_i(\vec{v}^*_{-i},\vec{x}^*,z^*)$, from which we deduce that $p_i(\vec{v}^*,\vec{x}^*,z^*)=0$. The payment \eqref{eq:pay} is thus derived by the envelop formula.
Following Myerson's result \citep{myerson1981optimal}, the COAD mechanism in Algorithm \ref{alg:coad} has the IC and IR properties, which completes the proof of Theorem \ref{prop:incentive}.
\end{proof}

\subsection{Proof of Theorem \ref{pp2}}\label{A1.3.1}
\begin{proof}~
\noindent
For any integer $m^*\geq 1$, to establish the desired result, it suffices to demonstrate that,
    \begin{equation*}
    R_{m^*+1}^{{\text{COAD}}|\mathcal{D}}(F_{{v^*}, {x^*}|z^*=\tilde{z}})\geq R_{m^*}^{{\text{COAD}}|\mathcal{D}}(F_{{v^*}, {x^*}|z^*=\tilde{z}} ), \text{ almost surely}.
    \end{equation*}
 Subsequently, by employing recursion, the desired result can be established. 

For any $m^*$ bidders with valuation $\{v_1^*,\dots,v_{m^*}^*\}\in\R^{m^*}_{\geq 0}$, by using the COAD mechanism, we can calculate the \change{pseudo-virtual value} $ c_i(v_i^*,x_i^*,z^*)$ of each bidder.  If 
\begin{equation*}
    \max_{i\in [m^*]}( c_i(v_i^*,x_i^*,z^*))=0,
\end{equation*}
the seller keeps the item and gets the 0 revenue.

If $\max_{i\in [m^*]}( c_i(v_i^*,x_i^*,z^*))>0$, without generation, suppose that the first $k\in\mathbb{N}_{+}$ bidders among these $m^*$ bidders satisfy $v_i^*\geq \hat{v}_i^L$. We can rewrite the set $\{v_1^*,\dots,v_k^*,\hat{v}_1^L,\dots,\hat{v}_k^L\}$ as $\{a_1,\dots,a_{2k}\}$ and let $a_{(2k-1)}$ be the second largest value among $\{a_1,\dots,a_{2k}\}$. Then the COAD mechanism will sell the item with price $\max\{0,a_{(2k-1)}\}$. 

Now, for each new bidder joining the auction with a valuation of $v_{m^*+1}^*\in\R_{\geq 0}$, we can implement the COAD mechanism for the $m^*+1$ bidders. 
\begin{itemize}
    \item If $v_{m^*+1}^*< \hat{v}^L_{m^*+1}$, then implementing the COAD mechanism for these $m^*+1$ bidders is equivalent to implementing the COAD mechanism for the first $m^*$ bidders, which means that the COAD mechanism will also sell the item at price $\max\{0,a_{(2k-1)}\}$ if $\max_{i\in [m^*]}( c_i(v_i^*,x_i^*,z^*))> 0$, otherwise keep the item and get 0 revenue.
    
    \item If $v_{m^*+1}^*\geq \hat{v}^L_{m^*+1}$, then we define  $a_{2k+1}$ as $v^*_{m^*+1}$ and $a_{2k+2}$ as $\hat{v}^L_{m^*+1}$, and let $a_{(2k+1)}$ denote the second highest value among $\{a_1,\dots,a_{2k},a_{2k+1},a_{2k+2}\}$. In such a scenario,  the COAD mechanism will sell the item with price $\max\{0,a_{(2k+1)}\}$. It follows that  $a_{(2k+1)}\geq a_{(2k-1)}$ for any $k\geq 1$, given that the second highest value in the set $\{a_1,\dots,a_{2k}\}$ will not exceed the second highest value in the set  $\{a_1,\dots,a_{2k},a_{2k+1},a_{2k+2}\}$.
\end{itemize}

 Based on the two different cases, conditioning on any first $m^*$ bidders, the revenue generated by the COAD mechanism on $m^*+1$ bidders will be no less than that achieved with  $m^*$ bidders for any additional bidder $m^*+1$, thus we have that,
 \begin{equation*}
    R_{m^*+1}^{{\text{COAD}}|\mathcal{D}}(F_{{v^*}, {x^*}|z^*=\tilde{z}})\geq R_{m^*}^{{\text{COAD}}|\mathcal{D}}(F_{{v^*}, {x^*}|z^*=\tilde{z}}), \text{ almost surely}.
\end{equation*}
Subsequently, employing recursion, we finalize the proof of Theorem \ref{pp2}.
\end{proof}

\subsection{Proof of Theorem \ref{prop1}}\label{A1.2}
\begin{proof}
\change{Let $\omega\in[m^*]$ be the bidder has the largest value, that is, $v^*_{\omega}=\max_{1\leq i\leq m^*} v_i^* $. If $v^*_{\omega} \geq \hat{v}^L_{\omega}$, then under COAD mechanism, the total payments satisfy
\begin{equation*}
    \sum^{m^*}_{i=1}p_i(\vec{v}^*,\vec{x}^*,z^*)\geq \max_{i\neq \omega}\{\hat{v}^L_{\omega}, v^*_i\}\geq \hat{v}^L_{\omega}.
\end{equation*}
Thus, it holds that $\sum^{m^*}_{i=1}p_i(\vec{v}^*,\vec{x}^*,z^*)\geq \hat{v}^L_{\omega}\mathbb{I}\{v^*_{\omega} \geq \hat{v}^L_{\omega}\}$.

By equation \eqref{rm},  it is defined that,
\begin{equation}\label{jasa 33}
\begin{split}
    R_{m^*}^{\mathcal M|\mathcal{D}}(F_{{v^*}, {x^*}|z^*=\tilde{z}})&= \mathbb{E}\left[\sum^{m^*}_{i=1} p_i(\vec{v}^*,\vec{x}^*,z^*) \big| \mathcal{D}, z^*=\tilde{z}\right]\\
    &\geq \mathbb{E}\left[\hat{v}^L_{\omega}\mathbb{I}\{v^*_{\omega} \geq \hat{v}^L_{\omega}\} \big| \mathcal{D}, z^*=\tilde{z}\right]\\
    &=\E\left[\hat{v}^L_{\omega}\mathbb{I}\{v^*_{\omega} \geq \hat{v}^L_{\omega}\} \big| v^*_{\omega} \geq \hat{v}^L_{\omega}, \mathcal{D}, z^*=\tilde{z}\right]\P(v^*_{\omega} \geq \hat{v}^L_{\omega} \big| \mathcal{D}, z^*=\tilde{z})\\
    &\quad+\E\left[\hat{v}^L_{\omega}\mathbb{I}\{v^*_{\omega} \geq \hat{v}^L_{\omega}\} \big| v^*_{\omega} < \hat{v}^L_{\omega}, \mathcal{D}, z^*=\tilde{z}\right]\P(v^*_{\omega} < \hat{v}^L_{\omega} \big| \mathcal{D}, z^*=\tilde{z})\\
    &=\E\left[\hat{v}^L_{\omega}\big| v^*_{\omega} \geq \hat{v}^L_{\omega}, \mathcal{D}, z^*=\tilde{z}\right]\P(v^*_{\omega} \geq \hat{v}^L_{\omega} \big| \mathcal{D}, z^*=\tilde{z}).
\end{split}
\end{equation}

By the definition of $\hat{v}^L_{\omega}$, we have that,
\begin{equation}\label{jasa 36}
\begin{split}
   \E\left[\hat{v}^L_{\omega}\big| v^*_{\omega} \geq \hat{v}^L_{\omega}, \mathcal{D}, z^*=\tilde{z}\right]&=\E\left[\hat{\mu}_n(x^*_{\omega},z^*)-S^*\big| v^*_{\omega} \geq \hat{v}^L_{\omega}, \mathcal{D}, z^*=\tilde{z}\right]\\
&=\E\left[{\mu}_n(x^*_{\omega},z^*)+\Delta_n(x^*_{\omega},z^*)-S^*\big| v^*_{\omega} \geq \hat{v}^L_{\omega}, \mathcal{D}, z^*=\tilde{z}\right],
\end{split}
\end{equation}
where $\Delta_n(x^*_{\omega},z^*)=\hat{\mu}_n(x^*_{\omega},z^*)-{\mu}_n(x^*_{\omega},z^*)$.

Based on Assumption \ref{assump:consistency} and Jensen's equality, we derive that,
\begin{equation*}
   |\E[\Delta_n(x_i^*,z^*)]|\leq  \E[|\Delta_n(x_i^*,z^*)|]\leq \sqrt{\E[\Delta_n^2(x_i^*,z^*)]}=O(n^{-\tau}).
\end{equation*}
Consequently, by employing Markov's inequality, we obtain that for any $i\in[m^*]$,
\begin{equation*}
    \E[|\Delta_n(x_i^*,z^*)|\ \big|\ v^*_{i} \geq \hat{v}^L_{i}, \mathcal{D}, z^*=\tilde{z}]=O_{\P}(n^{-\tau}).
\end{equation*}
Since $\omega$ is a random variable in $[m^*]$, using conditional probability, we can obtain
\begin{equation}\label{jasa 37}
\begin{split}
    &\E[|\Delta_n(x_{\omega}^*,z^*)|\ \big|\ v^*_{\omega} \geq \hat{v}^L_{\omega}, \mathcal{D}, z^*=\tilde{z}]\\
    =&\sum_{i=1}^{m^*}\E[|\Delta_n(x_{\omega}^*,z^*)|\ \big|\ v^*_{\omega} \geq \hat{v}^L_{\omega}, \mathcal{D}, z^*=\tilde{z}, \omega=i]\P(\omega=i)\\
    =&\sum_{i=1}^{m^*}\E[|\Delta_n(x_{i}^*,z^*)|\ \big|\ v^*_{i} \geq \hat{v}^L_{i}, \mathcal{D}, z^*=\tilde{z}]\P(\omega=i)\\
    =&O_{\P}(n^{-\tau})\sum_{i=1}^{m^*}\P(\omega=i)\\
    =&O_{\P}(n^{-\tau}).
\end{split}
\end{equation}

By the results in Section \ref{A.detail},  $S^*$ is the largest value such that $\eta^S_{1,n+1}<1-\alpha$. Consequently,  
given  $\mathcal{D}=\{D_{\text{train}}, D_{\text{cal}}\}$ and $z^*=\tilde{z}$, $S^*$ is determined accordingly. From Assumption \ref{assump:indepnoise}, $|\varepsilon^*_{\omega}|\leq C$ always holds. 
Thus, from \eqref{jasa 36}--\eqref{jasa 37} and the definition of $v^*_{\omega}$, we have that,
\begin{equation}\label{jasa 38}
    \begin{split}
         &\E\left[\hat{v}^L_{\omega}\big| v^*_{\omega} \geq \hat{v}^L_{\omega}, \mathcal{D}, z^*=\tilde{z}\right]\\
         =&\E[\mu_n(x^*_{\omega},z^*)\big| v^*_{\omega} \geq \hat{v}^L_{\omega}, \mathcal{D}, z^*=\tilde{z}]-S^*+O_{\P}(n^{-\tau})\\
         =& \E[\mu_n(x^*_{\omega},z^*)+\varepsilon^*_{\omega}\big| v^*_{\omega} \geq \hat{v}^L_{\omega}, \mathcal{D}, z^*=\tilde{z}]-\E[\varepsilon^*_{\omega}\big| v^*_{\omega} \geq \hat{v}^L_{\omega}, \mathcal{D}, z^*=\tilde{z}]-S^*+O_{\P}(n^{-\tau})\\   =&\E[v^*_{\omega}\big| v^*_{\omega} \geq \hat{v}^L_{\omega}, \mathcal{D}, z^*=\tilde{z}]-\E[\varepsilon^*_{\omega}\big| v^*_{\omega} \geq \hat{v}^L_{\omega}, \mathcal{D}, z^*=\tilde{z}]-S^*+O_{\P}(n^{-\tau})\\
         \geq& \E[v^*_{\omega}\big| v^*_{\omega} \geq \hat{v}^L_{\omega}, \mathcal{D}, z^*=\tilde{z}]-C-S^*+O_{\P}(n^{-\tau})\\
         \geq& \E[v^*_{\omega}\big|  z^*=\tilde{z}]-C-S^*+O_{\P}(n^{-\tau}).
    \end{split}
\end{equation}
The last inequality holds because conditioning on $ v^*_{\omega} \geq \hat{v}^L_{\omega}$ can only increase the expected value of 
 $v^*_{\omega}$, and $v^*_{\omega}$ is independent of the dataset $\mathcal{D}$.

It is known that \citep[see,][]{gibbs2023conformal}, under Assumptions \ref{assump:iiddata} and \ref{assump:indepbidders}, for any $i\in[m^*]$,
\begin{equation*}
\Big|\P(v_{i}^*\in [\hat{v}^L_{i},\hat{v}^U_{i}] \ |\ \mathcal{D}, z^*=\tilde{z})-(1-\alpha) \Big|=O_{\P}\left(\sqrt{\frac{q}{n}}\right).
\end{equation*}
Hence, for any $i\in[m^*]$,
\begin{equation}\label{jasa 34}
    \P(v^*_{i} \geq \hat{v}^L_{i} \big| \mathcal{D}, z^*=\tilde{z}) \geq 1-\alpha+O_{\P}(n^{-1/2}).
\end{equation}

Since $\omega$ is a random variable in $[m^*]$, using the conditional probability and \eqref{jasa 34}, we have that,
\begin{equation}\label{jasa 35}
\begin{split}
       \P(v^*_{\omega} \geq \hat{v}^L_{\omega} \big| \mathcal{D}, z^*=\tilde{z})&=\sum^{m^*}_{i=1}\P(v^*_{\omega} \geq \hat{v}^L_{\omega} \big| \mathcal{D}, z^*=\tilde{z},\omega=i)\P(\omega=i)\\
       &=\sum^{m^*}_{i=1}\P(v^*_{i} \geq \hat{v}^L_{i} \big| \mathcal{D}, z^*=\tilde{z})\P(\omega=i)\\
       &\geq (1-\alpha+O_{\P}(n^{-1/2}))\sum^{m^*}_{i=1}\P(\omega=i)\\
       &=1-\alpha+O_{\P}(n^{-1/2}).
\end{split}
\end{equation}

From \eqref{jasa 33}, \eqref{jasa 38}, and \eqref{jasa 35}, we can obtain 
\begin{equation*}
\begin{split}
    R_{m^*}^{COAD|\mathcal{D}}(F_{{v^*}, {x^*}|z^*=\tilde{z}}) &\geq (\E[v^*_{\omega}\big|   z^*=\tilde{z}]-C-S^*+O_{\P}(n^{-\tau}))(1-\alpha+O_{\P}(n^{-1/2}))\\
    &=(\E[v^*_{\omega}\big|  z^*=\tilde{z}]-C-S^*)(1-\alpha)+O_{\P}\left(n^{-\min\{\tau,1/2\}}\right)\\
    &=(1-\alpha)\left(\mathbb{E}_{}\left[\max_{1\leq i\leq m^*} v_i^*  \big|  z^*=\tilde{z}\right]-C-S^*\right)+O_{\P}\left(n^{-\min\{\tau,1/2\}}\right)\\
    &=(1-\alpha)(1-\lambda)W_{m^*}(F_{{v^*}|z^*=\tilde{z}})+O_{\P}\left(n^{-\min\{\tau,1/2\}}\right),
\end{split}
\end{equation*}
where $\lambda=(C+S^*)/W_{m^*}(F_{{v^*}|z^*=\tilde{z}})$.

This completes the proof of Theorem \ref{prop1}.}
\end{proof}

\subsection{Proof of Corollary \ref{cor 1}}\label{proof:cor1}
\begin{proof}
\change{From Proposition \ref{prop:length}, it follows that,  
\begin{equation*}
   0 < S^* \leq \max_{1 \leq j \leq n} |\Delta_n(x_j, z_j)| + C \leq \sup_{x_1,z_1}|\Delta_n(x_1,z_1)| +C.
\end{equation*}

In the worst-case scenario where \( S^* = \sup_{x_1, z_1} |\Delta_n(x_1, z_1)| + C \), it follows that  
\begin{equation*}
    \P(v \in [\hat{\mu}_n(x, z) - S^*, \hat{\mu}_n(x, z) + S^*]) = 1.
\end{equation*}  
Thus, we have that,
\begin{equation}\label{j 31}
    \P(v^*_{i} \geq \hat{v}^L_{i} \big| \mathcal{D}, z^*=\tilde{z}) =1.
\end{equation}

In the proof of Theorem \ref{prop1}, substituting \( S^* \) with \( \sup_{x_1, z_1} |\Delta_n(x_1, z_1)| + C \) in \eqref{jasa 38} and replacing \eqref{jasa 34} with \eqref{j 31}, we obtain  
\begin{equation*}
    R_{m^*}^{\text{COAD}|\mathcal{D}}(F_{{v^*}, {x^*}|z^*=\tilde{z}}) \geq W_{m^*}(F_{{v^*}|z^*=\tilde{z}}) - \left( \sup_{x_1,z_1}|\Delta_n(x_1,z_1)| + 2C \right) + O_{\P}\left(n^{-\tau}\right).
\end{equation*}

This completes the proof of Corollary \ref{cor 1}.}
\end{proof}

\subsection{Proof of Corollary \ref{theroem 6}}\label{proof:cor2}
\begin{proof}
\change{If condition 
\begin{equation*}  
\sup_{x_1,z_1}|\Delta_n(x_1,z_1)|+2C <  \left( 1 - \frac{1}{2\zeta \log_{\zeta} h} \right) h
\end{equation*} 
holds, it follows that
    \begin{equation*}
        \left(\sup_{x_1,z_1}|\Delta_n(x_1,z_1)|+2C\right) \Big/ \left(1-\frac{1}{2\zeta log_{\zeta}h}\right) < h.
    \end{equation*}

For any $\xi$ satisfying
    \begin{equation}\label{jasa 14}
           \left(\sup_{x_1,z_1}|\Delta_n(x_1,z_1)|+2C\right)  \Big/ \left(1-\frac{1}{2\zeta log_{\zeta}h}\right) <\xi <     h,
    \end{equation}
there exists an integer $M\in \mathbb{N}^+$ such that when $m^*>M$, 
\begin{equation}\label{jasa 15}
    \int^h_1 F_{v^*|z^*=\tilde{z}}^{m^*}(v)dv \leq h - \xi,
\end{equation}
where $F_{v^*|z^*=\tilde{z}}$ denotes the cumulative distribution function of the bidders' values given $z^*=\tilde{z}$.

Let $f_{{v^*}|z^*=\tilde{z}}$ denote the density function of the bidders' values given $z^*=\tilde{z}$. From \eqref{jasa 15},
it follows that,
\begin{equation}\label{jasa 16}
\begin{split}
        W_{m^*}(F_{{v^*}|z^*=\tilde{z}}) &= \mathbb{E}_{}\left[\max_{1\leq i\leq m^*} v_i^*  \big|  z^*=\tilde{z}\right]\\
    &=\int^h_1vm^*F^{m^*-1}_{{v^*}|z^*=\tilde{z}}(v)f_{{v^*}|z^*=\tilde{z}}(v)dv\\
    &=\int^h_1 v dF^{m^*}_{{v^*}|z^*=\tilde{z}}(v)\\
    &=h- \int^h_1 F_{v^*|z^*=\tilde{z}}^{m^*}(v)dv\\
    &\geq\xi.
\end{split}
\end{equation}

Therefore, for $m^*>M$, by \eqref{jasa 14} and \eqref{jasa 16}, it follows that,
\begin{equation*}
      \left(\sup_{x_1,z_1}|\Delta_n(x_1,z_1)|+2C\right) \Big/\left(1-\frac{1}{2\zeta log_{\zeta}h}\right) < W_{m^*}(F_{{v^*}|z^*=\tilde{z}}).
\end{equation*}

Consequently, we obtain,
    \begin{equation}\label{jasa 18}
    \begin{split}
        &W_{m^*}(F_{{v^*}|z^*=\tilde{z}}) - \left( \sup_{x_1,z_1}|\Delta_n(x_1,z_1)| + 2C \right) \\
        >&W_{m^*}(F_{{v^*}|z^*=\tilde{z}})-\left(1-\frac{1}{2\zeta log_{\zeta}h}\right)W_{m^*}(F_{{v^*}|z^*=\tilde{z}})\\
        =&\frac{1}{2\zeta log_{\zeta}h}W_{m^*}(F_{{v^*}|z^*=\tilde{z}}).
    \end{split}
    \end{equation}

By substituting \eqref{jasa 18} into Corollary \ref{cor 1}, we obtain
 \begin{equation*}
    R_{m^*}^{{\text{COAD}}|\mathcal{D}}(F_{{v^*}, {x^*}|z^*=\tilde{z}})\geq \frac{1}{2\zeta log_{\zeta}h}W_{m^*}(F_{{v^*}|z^*=\tilde{z}})+O_{\P}\left(n^{-\tau}\right).
\end{equation*}    

This completes the proof of Corollary \ref{theroem 6}.}
\end{proof}

\section{Additional Numerical Results}\label{app:add_exp}
\change{In this section, we present additional simulation results to evaluate the performance of our method when the key assumptions of the paper are \emph{not} satisfied and when the dataset is larger, involving more bidders and items.
Following the setup in Section \ref{sec:four}, we compare the expected revenue generated by COAD against two benchmarks: the second-price auction and Myerson's auction based on the empirical distribution.}
We set the miscoverage level at $\alpha=0.1$. In each simulation, we randomly generate $N$ iid data points based on the regression model. We use half of the data as calibration data and the remaining half as training data. Error bars in the line plots indicate the 95\% CI for the mean.

\subsection{{Results under Violated Assumptions}}\label{low_d}

\change{We consider an auction setting where several key assumptions of our framework are \emph{not} satisfied.
For each data point \((x_j, z_j, v_j) \in \mathcal{D}\), the item feature \( z_j \) is uniformly drawn from the set \( \mathcal{Z} = \{3,5,7\} \). The bidder feature \( x_j \) follows an autoregressive process:
\[
x_j = \rho_x x_{j-1} + \sqrt{1 - \rho_x^2} \xi_j, \quad j = 1, \dots, N,
\]
with \( x_0 = 0 \), \( \rho_x = 0.2 \), and \( \xi_j \sim \mathcal{N}(z_j/10,1) \). 
This temporal dependence violates Assumption \ref{assump:iiddata}, which requires that the historical data be independent and identically distributed.
We define the value approximation error as $v_j - \mu(x_j, z_j) = (x_j^2 \cos z_j) \cdot \eta_j$, 
where \( \eta_j \) is independently drawn from a truncated standard normal distribution on \([-1,1]\).
This violates Assumption \ref{assump:indepnoise}, which assumes the error is bounded.
The previous empirical findings of \citet{ostrovsky2023reserve} and \citet{lahaie2007revenue} suggest that in practice, the values of the bidders for a given item are drawn from log-normal distributions. Similarly, we assume that for any $(x,z)\in\mathcal{X}\times \mathcal{Z}$, the regression model is
\begin{equation*}
    \mu(x,z)=e^{x}z. 
\end{equation*}}
\change{For a given item with feature \( z_j \), the function \( \mu(x_j, z_j) \) follows a log-normal distribution, as \( x_j \) is normally distributed.}
This setup represents a simplified eBay online auction, where three types of items are sold. Each bidder has a continuous feature, which can represent an infinite variety of bidders even with just one dimension. A correlation is introduced between item and bidder features, as different items are expected to attract distinct groups of bidders. 
We fit the model using eighth-order polynomial regression.

\change{For a new auction involving an item with feature $z^*\in\mathcal{Z}$ and $m^*$ participating bidders, each bidder’s feature $x_i^*$, for $i\in[m^*]$, is generated using an autoregressive process:
\[
x_i^* = \rho_x x^*_{i-1} + \sqrt{1 - \rho_x^2} \xi_i^*, \quad i = 1, \dots, m^*,
\]
with \( x_0^* = 0 \), \( \rho_x = 0.2 \), and \( \xi_j^* \sim \mathcal{N}(z^*/10,1) \). The value approximation error is defined as $v_i^* - \mu(x_i^*, z_i^*) = ({x_i^*}^2 \cos z_i^*) \cdot \eta_i^*$, where \( \eta_i^* \) is drawn from a truncated standard normal distribution on \([-1,1]\). This setup introduces dependence across bidders through the autoregressive structure, thereby violating Assumption \ref{assump:indepbidders}, which requires independence among new bidders.
Combined with the violations of data independence and error assumptions in the training data, this simulation setting violates Assumptions \ref{assump:iiddata} through \ref{assump:indepbidders}.}

\begin{figure}[ht!]
    \centering
    \begin{subfigure}[b]{0.29\textwidth}
        \centering
        \includegraphics[width=0.8\textwidth]{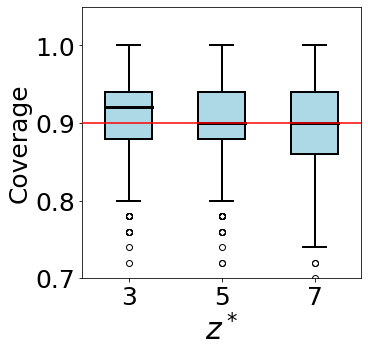}
        \caption{}
        \label{fig:5a}
    \end{subfigure}
        \begin{subfigure}[b]{0.67\textwidth} 
        \centering
        \includegraphics[width=\textwidth]{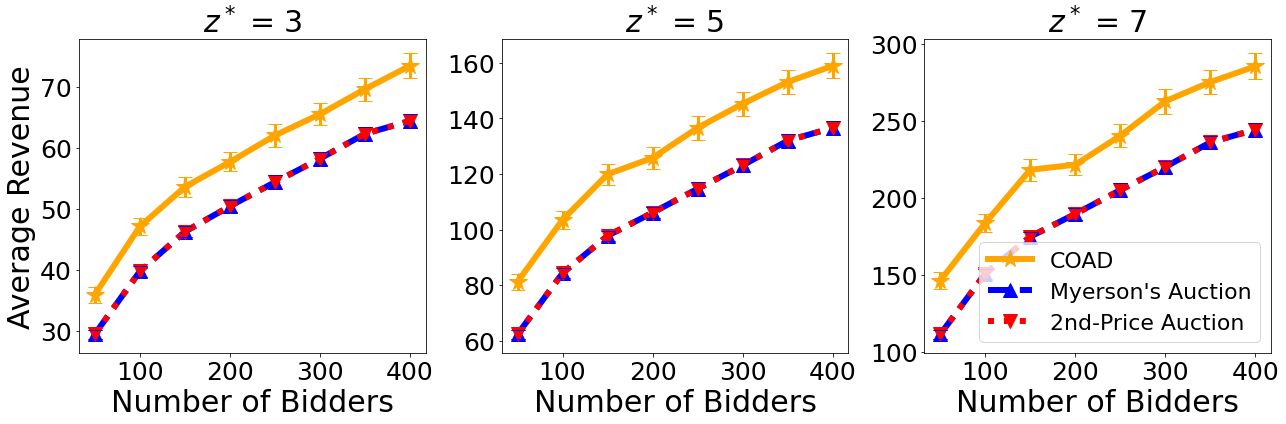}
        \caption{}
        \label{fig:5b}
    \end{subfigure}
    \begin{subfigure}[b]{0.67\textwidth}
        \centering
        \includegraphics[width=\textwidth]{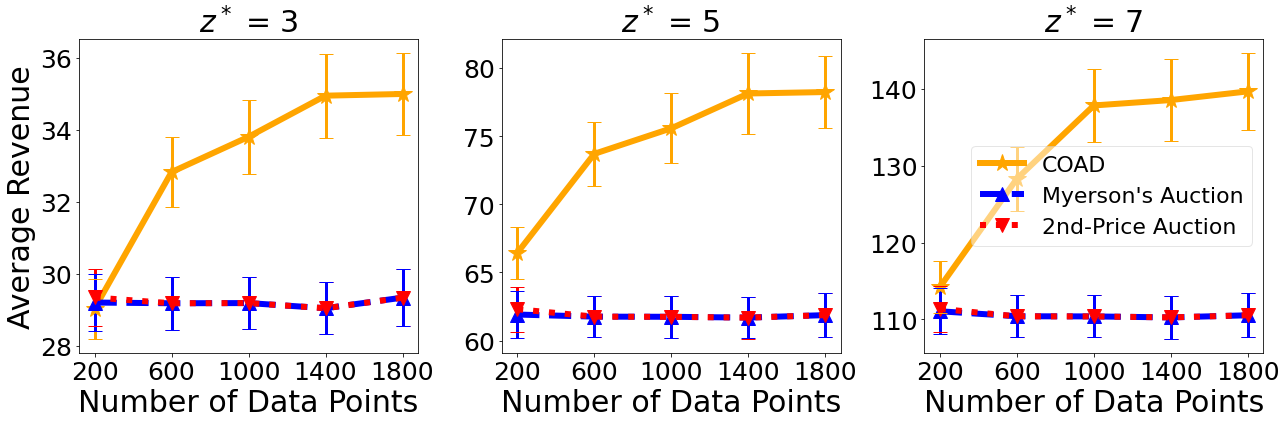}
        \caption{}
        \label{fig:5c}
    \end{subfigure}
    \caption{\small{\change{Results from Section \ref{low_d}, based on 1000 experiments. (a) Boxplots of the coverage probability for the true value, conditioned on various $z^*$ with $N=1000$ and $m^*=50$. The red line indicates the target level of $1-\alpha=0.9$. (b) Average revenue of different mechanisms for various features $z^*$, with varying numbers of $m^*\in\{50,100,150,200,250,300,350,400\}$ and $N=5000$. (c)  Average revenue of different mechanisms for various features $z^*$, with varying numbers of $N\in\{200, 600, 1000, 1400, 1800\}$ and $m^*=50$.} }}
    \label{fig:combined1}
\end{figure}

\change{Figure \ref{fig:5a} presents boxplots of the conditional coverage probabilities for the conformal prediction intervals, conditioned on different item features. Specifically, it shows the empirical distribution of $\P(v_i^*\in \hat{C}_{dual}(x_i^*,z^*)\ |\ \mathcal{D}, z^*=\tilde{z})$ for various $\tilde{z}\in \mathcal{Z}$. The results indicate that the conformal prediction method achieves coverage close to the target level $1-\alpha$, even under assumption violations. This supports the practical reliability of the method introduced in Section \ref{sec:conformal}, demonstrating its robustness when Assumptions \ref{assump:iiddata} and \ref{assump:indepbidders} are not fully satisfied.
Figure \ref{fig:5b} shows the average revenue of different mechanisms across auctions with various item features and varying numbers of bidders.  As \( m^* \) increases, the expected revenue of COAD  also increases, consistent with Theorem \ref{pp2}. In comparison, Myerson’s auction with a single reserve price and the standard second-price auction yield similar revenue, especially when the number of bidders is large. This aligns with prior findings that the impact of a single reserve price is significant only when the number of bidders is small  \citep{ostrovsky2023reserve}.
Figure \ref{fig:5c} presents the average revenue of the different mechanisms across auctions with various item features and increasing dataset sizes. Across all scenarios, COAD consistently outperforms both the empirical Myerson auction and the second-price auction. Moreover, as the dataset size grows, the average revenue of COAD improves. This is due to the reduction in prediction error with more training data, which enhances reserve price accuracy and ultimately boosts revenue—consistent with the theoretical insights in Section \ref{sec:sensitivity}. Overall, these results highlight the strong empirical performance of COAD, even when the main theoretical assumptions are violated.}

\subsection{Results with A Larger Number of Bidders and Items}\label{larger}

\change{We consider an auction setting similar to that in Section \ref{sec:highdsimnn}, but with a larger number of bidders and item types. Specifically, both bidder and item features are $20$-dimensional, i.e.,  $x\in\R^{20}, z\in\mathcal{Z}\subset\R^{20}$. Each item feature $z$ is uniformly sampled from a larger set $\mathcal{Z}=\{\tilde{z}_1,\dots,\tilde{z}_{q}\}$ with $q=50$, where each $\tilde{z}_{i}$ is drawn independently from $\mathcal{N}(\vec{\boldsymbol{0}}, \boldsymbol{I}_{20})$. The bidder feature $x$ is generated from $\mathcal{N}(\boldsymbol{\mu_x}, \boldsymbol{I}_{20})$, where $\boldsymbol{\mu_x}=(||z||_2^2/{20}, ||z||_2^2/{20},\dots,||z||_2^2/{20})$.
The regression model is,
\begin{equation*}
    \mu(x,z) = e^{\beta_1^\top  x}\cdot (\beta_2^\top  z),
\end{equation*}
where  $\beta_1\in\mathbb{R}^{20}, \beta_2\in\mathbb{R}^{20}$, with each entry of $\beta_1$ and $\beta_2$ independently drawn from the uniform distribution Unif$[-0.5,0.5]$. This formulation is motivated by prior empirical studies \citep{ostrovsky2023reserve, lahaie2007revenue} showing that bidder values for a given item often follow a log-normal distribution.
To model value approximation error, we set \( v - \mu(x, z) = e^{\cos^2(x^\top z)} \eta \), where \( \eta \) is independently drawn from \(\text{Unif}[-1,1]\).

This setup simulates a realistic online advertising environment where ad slots are associated with $50$ different keyword types. The item features capture information about the keywords, while bidder features represent attributes of the advertisements, such as the advertiser's rating, brand identity, and product information.  To fit the model, we use a neural network with two hidden layers containing 128 and 64 neurons, respectively. The network uses LeakyReLU activation functions \citep{maas2013rectifier}, incorporates $L_2$ regularization, applies  $30\%$ dropout, and is trained using the Adam optimizer. Training is conducted for $15$ epochs with a batch size of $32$.}

\begin{figure}[t!]
    \centering
    \begin{subfigure}[b]{0.267\textwidth}
        \centering
        \includegraphics[width=\textwidth]{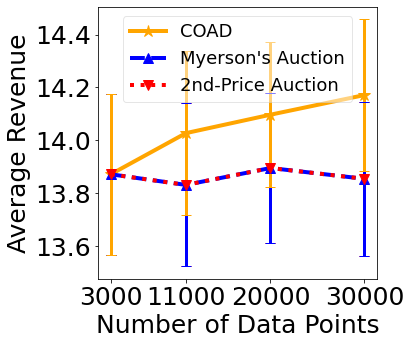}
        \caption{}
        \label{fig:nn-a_app}
    \end{subfigure}
    \begin{subfigure}[b]{0.243\textwidth}
        \centering
        \includegraphics[width=\textwidth]{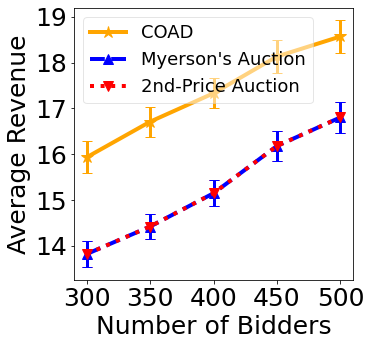}
        \caption{}
        \label{fig:nn-b_app}
    \end{subfigure}
    \begin{subfigure}[b]{0.25\textwidth}
        \centering
        \includegraphics[width=\textwidth]{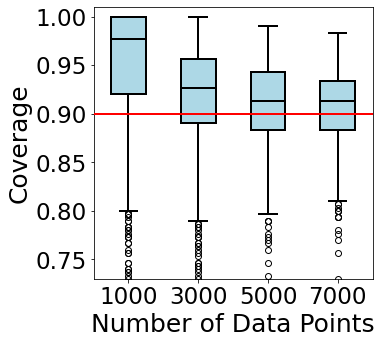}
        \caption{}
        \label{fig:nn-c_app}
    \end{subfigure}
    \caption{\small{\change{Results from Section \ref{larger}, based on 1000 experiments conducted for a randomly selected item.  (a) Average revenue of different mechanisms, with varying numbers of $N\in\{3000,11000, 20000, 30000\}$ and $m^*=300$. (b) Average revenue of different mechanisms, with varying numbers of $m^*\in \{300, 350, 400, 450, 500\}$ and $N=50000$. (c) Boxplots of the coverage probability for the true value, with varying numbers of $N\in\{1000,3000,5000,7000\}$ and $m^*=300$. The red line indicates the target level \( 1 - \alpha = 0.9 \).} }}
    \label{fig:combined:high2_app}
\end{figure}

\change{Figure \ref{fig:nn-a_app} shows the average revenue of the three auction mechanisms across different dataset sizes, with the number of bidders fixed at \( m^* = 300 \). COAD consistently outperforms both the second-price auction and the empirical Myerson auction in terms of average revenue. Its performance improves as more data becomes available—consistent with the theoretical insights presented in Section \ref{sec:sensitivity}.
Figure \ref{fig:nn-b_app} illustrates the average revenue of the three mechanisms as the number of bidders increases, with \( m^* \in \{300, 350, 400, 450, 500\} \). The expected revenue of COAD increases with the number of bidders $m^*$, aligning with Theorem \ref{pp2}. Notably, the empirical Myerson auction with a single reserve price performs similarly to the second-price auction in this high-bidder regime, as the effect of a single reserve price diminishes when the bidder pool is large \citep{ostrovsky2023reserve}.
Figure \ref{fig:nn-c_app} presents boxplots of the conditional coverage of the conformal prediction intervals across different dataset sizes. The plots show the empirical distribution of \( \P(v_i^* \in \hat{\mathcal{C}}_{\text{dual}}(x_i^*,z^*) \mid \mathcal{D}, z^*=\tilde{z}) \). As the data size increases, the coverage becomes more concentrated around the target level \( 1 - \alpha \),  confirming that the method described in Section \ref{sec:propconformal} achieves the desired conditional coverage. Overall, COAD scales effectively with a large number of bidders and item types.
}

\end{document}